\documentclass[12pt,a4paper]{scrartcl}

\usepackage{makeidx}
\usepackage{comment}
\usepackage{algorithm}
\usepackage{algorithmicx}
\usepackage[noend]{algpseudocode}
\usepackage{url}
\usepackage{graphicx}
\usepackage{multirow}
\usepackage{amsmath}
\usepackage{latexsym}

\newtheorem{Theorem}{Theorem}

\newtheorem{prob}{Problem}

\newenvironment{Proof}{\trivlist \item[\hskip \labelsep{\bf proof.\/}]}{\hspace{\fill}$\Box$\endtrivlist }

\newcommand{\logstar}{\lg^*\hspace{-.9mm}}

\begin{document}

\title{siEDM: an efficient string index and search algorithm for edit distance with moves}
\author{Yoshimasa Takabatake$^{1}$, Kenta Nakashima$^{1}$, Tetsuji Kuboyama$^{2}$, \\Yasuo Tabei$^{3}$ and Hiroshi Sakamoto$^{1}$\thanks{mailto: hiroshi@ai.kyutech.ac.jp}\\
{\small $^{1}$ Kyushu Institute of Technology, Japan}\\
{\small $^{2}$ Gakushuin University, Japan}\\
{\small $^{3}$ PRESTO, Japan Science and Technology Agency, Japan}
}
\date{\empty}
\maketitle

    \abstract{ Although several self-indexes for highly repetitive text
      collections exist, developing an index and search algorithm with editing
      operations remains a challenge.  \emph{Edit distance with moves (EDM)}
      is a string-to-string distance measure that includes substring moves in
      addition to ordinal editing operations to turn one string into another.
      Although the problem of computing EDM is intractable, it has a wide
      range of potential applications, especially in approximate string
      retrieval.
      Despite the importance of computing EDM, there has been no efficient
      method for indexing and searching large text collections based on the
      EDM measure.  We propose the first algorithm, named \emph{string index
        for edit distance with moves (siEDM)}, for indexing and searching
      strings with EDM.
      The siEDM algorithm builds an index structure by leveraging the idea
      behind the \emph{edit sensitive parsing (ESP)}, an efficient algorithm
      enabling approximately computing EDM with guarantees of upper and lower
      bounds for the exact EDM.
      siEDM efficiently prunes the space for searching query strings
      by the proposed method, which enables fast query searches with
      the same guarantee as ESP.  We experimentally tested the ability
      of siEDM to index and search strings on benchmark datasets, and
      we showed siEDM's efficiency.  }

\section{Introduction}

The vast amounts of text data are created, replicated, and modified with the
increasing use of the internet and advances of data-centric technology. Many
of these data contain repetitions of long substrings with slight differences,
so called \emph{highly repetitive texts}, such as Wikipedia and software
repositories like GitHub with a large number of revisions.  Also recent
biological databases store a large amount of human genomes while the genetic
differences among individuals are less than $0.1$ percent, which results in
the collections of human genomes to be highly repetitive.  Therefore, there is
a strong need to develop powerful methods for processing highly repetitive
text collections on a large scale.

Building indexes is the \emph{de facto} standard method to search large
databases of highly repetitive texts.  Several methods have been presented for
indexing and searching large-scale and highly repetitive text collections.
Examples include the ESP-index~\cite{Takabatake14}, SLP-index~\cite{Claude09}
and LZ77-based index~\cite{GagieGKNP14}.  Recently, Gagie and
Puglisi~\cite{Gagie:2015bj} presented a general framework called kernelization
for indexing and searching highly repetitive texts.  Although these methods
enable fast query searches, their applicability is limited to exact match
searches.

The edit distance between two strings is the minimum cost of edit operations
(insertions, deletions, and replacements of characters) to transform one
string to another. It has been proposed for detecting evolutionary changes in
biological sequences~\cite{Durbin98}, detecting typing errors in
documents~\cite{Crochemore94}, and correcting errors on lossy communication
channels~\cite{Levenshtein96}.  To accelerate the quadratic time upper bound
on computing the edit distance, Cormode and Muthukrishnan introduced a new
technique called \emph{edit sensitive parsing (ESP)}~\cite{Cormode07}.  This
technique allows us to compute a modified edit distance in near linear time by
sacrificing accuracy with theoretical bounds.  The modified distance is known
as \emph{edit distance with moves (EDM)}~\cite{Cormode07}, which includes
substring move operations in addition to insertions and deletions.
While the exact computation of EDM is known to be intractable~\cite{Shapira07},
the approximate computation of EDM using ESP achieves a good approximation ratio
$O(\lg{N}\lg^*{N})$, and runs in almost linear time $O(N\lg^*{N})$ for the
string length $N$, where $\lg$ denotes the logarithm of base two.

ESP is extended to various applications for highly repetitive texts.  Examples
are data compressions called grammar compression~\cite{SakamotoMKS09,
  Maruyama12, Maruyama13, Maruyama14}, indexes for exact matches~\cite{ESP,
  Takabatake14, Takabatake15}, an approximated frequent pattern
discovery~\cite{Nakahara13} and an online pattern matching for
EDM~\cite{Takabatake14-2}.  Despite several attempts to efficiently compute
EDM and various extensions of ESP, there is no method for indexing and
searching texts with EDM.  Such a method is required in bioinformatics where
approximated text searches are used to analyze massive genome sequences.
Thus, an open challenge is to develop an efficient string index and search
algorithm for EDM.

We propose a novel method called siEDM that efficiently indexes massive text,
and performs query searches for EDM.  As far as we know, siEDM is the first
string index for searching queries for EDM.  A space-efficient index structure
for a string is built by succinctly encoding a parse tree obtained from ESP,
and query searches are performed on the encoded index structures.  siEDM
prunes useless portions of the search space based on the lower bound of
EDM without missing any matching patterns, enabling fast query searches.
As in existing methods, similarity searches of siEDM are approximate but have
the same guarantee of the approximation ratio as in ESP.

Experiments were performed on indexing and searching repetitive texts for EDM
on standard benchmark datasets.  The performance comparison with an online
pattern matching for EDM~\cite{Takabatake14-2} demonstrates siEDM's
practicality.

\section{Preliminaries}
\subsection{Basic notations}

Let $\Sigma$ be a finite alphabet, and $\sigma$ be $|\Sigma|$.  All elements
in $\Sigma$ are totally ordered.  Let us denote by $\Sigma^*$ the set of all
strings over $\Sigma$, and by $\Sigma^q$ the set of strings of length $q$ over
$\Sigma$, i.e., $\Sigma^q=\{w\in \Sigma^*:|w|=q\}$ and an element in
$\Sigma^q$ is called a $q$-gram.  The length of a string $S$ is denoted by
$|S|$.  The empty string $\epsilon$ is a string of length $0$, namely
$|\epsilon|=0$.  For a string $S=\alpha\beta\gamma$, $\alpha$, $\beta$ and
$\gamma$ are called the prefix, substring, and suffix of $S$, respectively.
The $i$-th character of a string $S$ is denoted by $S[i]$ for $i \in [1,|S|]$.
For a string $S$ and interval $[i,j]$ ($1 \leq i \leq j \leq |S|$), let
$S[i,j]$ denote the substring of $S$ that begins at position $i$ and ends at
position $j$, and let $S[i,j]$ be $\epsilon$ when $i>j$.
For a string $S$ and integer $q \geq 0$, let
$\mathit{pre}(S,q)=S[1,q]$ and $\mathit{suf}(S,q)=S[|S|-q+1,|S|]$.  
We assume a recursive enumerable set
${\cal X}$ of variables with $\Sigma\cap {\cal X}=\emptyset$.  All elements in
$\Sigma\cup {\cal X}$ are totally ordered, where all elements in $\Sigma$ must
be smaller than those in ${\cal X}$.  In this paper, we call a sequence of
symbols from $\Sigma \cup X$ a string.  Let us define $\lg^{(1)}{u}=\lg{u}$,
and $\lg^{(i+1)}{u}=\lg{(\lg^{(i)}{u})}$ for $i\geq 1$.  The iterated
logarithm of $u$ is denoted by $\lg^*{u}$, and defined as the number of times
the logarithm function must be applied before the result is less than or equal
to $1$, i.e., $\lg^*u=\min\{i:\lg^{(i)}{u} \leq 1\}$.

\subsection{Straight-line program (SLP)}

A context-free grammar (CFG) in Chomsky normal form is a quadruple
$G=(\Sigma,V,D,X_s)$, where $V$ is a finite subset of ${\cal X}$, $D$ is a
finite subset of $V\times (V\cup \Sigma)^2$, and $X_s \in V$ is the start
symbol.  An element in $D$ is called a production rule.  Denote $X_{l(k)}$
(resp. $X_{r(k)}$) as a left symbol (resp. right symbol) on the right hand
side for a production rule with a variable $X_k$ on the left hand side, i.e.,
$X_k\to X_{l(k)}X_{r(k)}$.  $\mathit{val}(X_i)$ for variable $X_i \in V$ denotes the
string derived from $X_i$.  A grammar compression of $S$ is a CFG $G$ that
derives $S$ and only $S$.  The size of a CFG is the number of variables, i.e.,
$|V|$ and let $n=|V|$.

The parse tree of $G$ is a rooted ordered binary tree such that (i) internal
nodes are labeled by variables in $V$ and (ii) leaves are labeled by symbols in
$\Sigma$, i.e., the label sequence in leaves is equal to the input string.  In
a parse tree, any internal node $Z$ corresponds to a production rule $Z\to XY$,
and has the left child with label $X$ and the right child with label $Y$.

\emph{Straight-line program (SLP)}~\cite{SLP} is defined as a grammar
compression over $\Sigma\cup V$, and its production rules are in the form of
$X_k\to X_iX_j$ where $X_i,X_j \in \Sigma\cup V$ and
$1\leq i,j<k\leq n + \sigma$.

\subsection{Rank/select dictionaries}

A rank/select dictionary for a bit string $B$~\cite{Jacobson89} supports the
following queries: $\mathit{rank}_c(B,i)$ returns the number of occurrences of
$c \in \{0,1\}$ in $B[0,i]$; $\mathit{select}_c(B,i)$ returns the position of
the $i$-th occurrence of $c\in\{0,1\}$ in $B$; $\mathit{access}(B,i)$ returns
the $i$-th bit in $B$.  Data structures with only the $|B| + o(|B|)$ bits
storage to achieve $O(1)$ time rank and select queries~\cite{Raman07} have
been presented.

GMR~\cite{Golynski06} is a rank/select dictionary for large alphabets and
supports rank/ select/access queries for strings in $(\Sigma\cup V)^*$.  GMR
uses $(n+\sigma)\lg{(n+\sigma)}+o((n+\sigma)\lg{(n+\sigma)})$ bits while
computing both rank and access queries in $O(\lg{\lg{(n+\sigma)}})$ times and
also computing select queries in $O(1)$ time.

\section{Problem}

We first review the notion of EDM.
The distance $d(S,Q)$ between two strings $S$ and $Q$ is 
the minimum number of edit operations to transform $S$ into $Q$.
The edit operations are defined as follows:
\begin{enumerate}
\item Insertion: A character $a$ is inserted at position $i$ in $S$, which generates $S[1,i-1]aS[i,|S|]$,
\item Deletion: A character is deleted at position $i$ in $S$, which generates $S[1,i-1]S[i+1,|S|]$,
\item Replacement: A character is replaced by $a$ at position $i$ in $S$, which generates $S[1,i-1]aS[i+1,|S|]$,
\item Substring move: A substring $S[i,j]$ is deleted from the position $i$,
  and inserted at the position $k$ in $S$, which generates
  $S[1,i-1]S[j+1,k-1]S[i,j]S[k,|S|]$ for $1\leq i\leq j\leq k\leq |S|$, and
  $S[1,k-1]S[i,j]S[k,i-1]S[j+1,|S|]$ for $1\leq k\leq i\leq j\leq |S|$.
\end{enumerate}

\begin{prob}[Query search for EDM] \label{prob:1}
For a string $S\in \Sigma^*$, a query $Q\in \Sigma^*$ and a distance threshold $\tau\geq 0$, find all $i\in [1,|S|]$ 
satisfying $d(S[i,i+|Q|-1],Q)\leq \tau$.
\end{prob}

Shapira and Storer~\cite{Shapira07} proved the NP-completeness of EDM and
proposed a polynomial-time algorithm for a restricted EDM.
Cormode and Muthukrishnan~\cite{Cormode07} presented an approximation
algorithm named ESP for computing EDM.  We present a string index and
search algorithm by leveraging the idea behind ESP for solving Problem~\ref{prob:1}.  
Our method consists of two parts: (i) an efficient index structure for a given string $S$ and (ii) a fast
algorithm for searching query $Q$ on the index structure of $S$ with
respect to EDM.  Although our method is also an approximation
algorithm, it guarantees upper and lower bounds for the exact EDM.  We
first review ESP in the next section and then discuss the two parts.

\section{Edit Sensitive Parsing~(ESP) for building SLPs}

\subsection{ESP revisit}

We review the edit sensitive parsing algorithm for building SLPs~\cite{SakamotoMKS09}.
This algorithm, referred to as ESP-comp, computes an SLP from an input sting $S$.
The tasks of ESP-comp are to (i) partition $S$ into $s_1s_2\cdots s_\ell$ such that
$2\leq |s_i|\leq 3$ for each $1\leq i\leq \ell$,
(ii) if $|s_i|=2$, generate the production rule $X\to s_i$ and replace $s_i$ by $X$
(this subtree is referred to as a 2-tree), and 
if $|s_i|=3$, generate the production rule $Y\to AX$ and $X\to BC$ for $s_i=ABC$,
and replace $s_i$ by $Y$ (referred to as a 2-2-tree),
(iii) iterate this process until $S$ becomes a symbol.
Finally, the ESP-comp builds an SLP representing the string $S$.

We focus on how to determine the partition $S=s_1s_2\cdots s_\ell$.
A string of the form $a^r$ with $a\in\Sigma\cup V$ and $r\geq 2$ is called a repetition.
First, $S$ is uniquely partitioned into the form
$w_1x_1w_2x_2\cdots w_kx_kw_{k+1}$  by its maximal repetitions, where each $x_i$ is 
a maximal repetition of a symbol in $\Sigma\cup V$, and each $w_i\in(\Sigma\cup V)^*$ contains no repetition.
Then, each $x_i$ is called type1, each $w_i$ of length at least $2\lg^*|S|$ is type2, and any remaining $w_i$ is type3.
If $|w_i|=1$, this symbol is attached to $x_{i-1}$ or $x_i$ with preference $x_{i-1}$ when both cases are possible.
Thus, if $|S|>2$, each $x_i$ and $w_i$ is longer than or equal to two.
One of the substrings is referred to as $S_i$.

Next, ESP-comp parses each $S_i$ depending on the type.
For type1 and type3 substrings, the algorithm performs the \emph{left aligned parsing} as follows. 
If $|S_i|$ is even, the algorithm builds 2-tree from $S_i[2j-1,2j]$ for each $j \in \{1,2,\ldots,|S_i|/2\}$;
otherwise, it builds a 2-tree from $S_i[2j-1,2j]$ for each $j \in \{1,2,\ldots,\lfloor(|S_i|-3)/2\rfloor\}$ and 
builds a 2-2-tree from the last trigram $S_i[|S_i|-2,|S_i|]$. 
For type2 $S_i$, the algorithm further partitions it into short substrings of length two or three by
\emph{alphabet reduction}~\cite{Cormode07}.

{\bf Alphabet reduction:} 
Given a type2 string $S$, consider $S[i]$ and $S[i-1]$ as binary integers.
Let $p$ be the position of the least significant bit, in which $S[i]\neq S[i-1]$,
and let $bit(p,S[i])$ be the bit of $S[i]$ at the $p$-th position.
Then, $L[i]=2p+bit(p,S[i])$ is defined for any $i\geq 2$.
Because $S$ is repetition-free (i.e., type2), the label string $L(S)=L[2]L[3]\cdots L[|S|]$ is also type2.
If the number of different symbols in $S$ is $n$ (denoted by $[S]$), then $[L(S)]=O(\lg n)$.
For the $L(S)$, the next label string is iteratively computed until the final $L^*(S)$ satisfying
$[L^*(S)]\leq \lg^*|S|$ is obtained.
$S[i]$ is called the \emph{landmark} if $L[i]> \max\{L[i-1],L[i+1]\}$. 

The alphabet reduction transforms $S$ into $L^*(S)$ such that any substring of $L^*(S)$ of 
length at least $2\lg^*|S|$ contains at least one landmark because $L^*(S)$ is also type2.
Using this characteristic, the algorithm ESP-comp determines the bigrams $S[i,i+1]$ to be replaced
for any landmark $S[i]$, where any two landmarks are not adjacent, and then the replacement is deterministic.
After replacing all landmarks, any remaining maximal substring $s$ is replaced by the left aligned parsing,
where if $|s|$ =1, it is attached to its left or right block.

\begin{figure}[t]
\begin{center}
\includegraphics[width=0.8\textwidth]{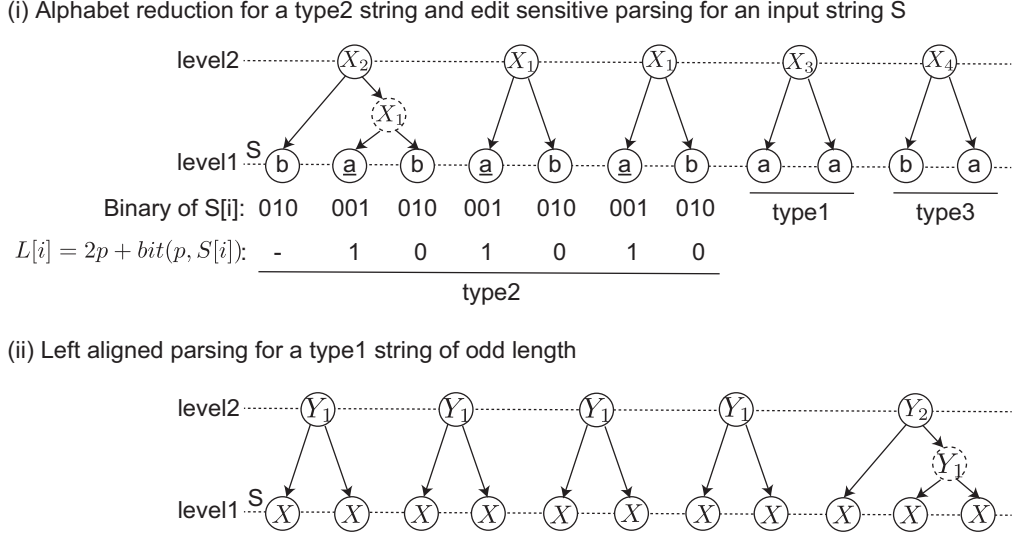}
\end{center}
\vspace{0.5cm}
\caption{The edit sensitive parsing. In (i), an underlined $S[i]$ means a landmark,
and $p\geq 0$.
In (i) and (ii), a dashed node is corresponding to the intermediate node in a 2-2-tree.}
\label{fig:reduction}
\end{figure}

We give an example of the edit sensitive parsing of an input string in Figure~\ref{fig:reduction}-(i) and (ii).
The input string $S$ is divided into three maximal substrings depending on the types.
The label string $L$ is computed for the type2 string.
Originally, $L$ is iteratively computed until $[L]\leq \lg^*|S|$.
This case shows that a single iteration satisfies this condition.
After the alphabet reduction, three landmarks $S[i]$ are found, and then each $S[i,i+1]$ is parsed.
Any other remaining substrings including type1 and type3 are parsed by the left aligned parsing
shown in Figure~\ref{fig:reduction}-(ii).
In this example, a dashed node denotes that it is an intermediate node in a 2-2-tree.
Originally, an ESP tree is a ternary tree in which each node has at most three children.
The intermediate node is introduced to represent ESP tree as a binary tree.

As shown in~\cite{Cormode07}, the alphabet reduction approximates the
minimum CFG as follows.  Let $S$ be a type2 string containing a
substring $\alpha$ at least twice.  When $\alpha$ is sufficiently long
(e.g., $|\alpha|\geq 2\lg^*|S|$), there is a partition
$\alpha = \alpha_1\alpha_2$ such that $|\alpha_1|=O(\lg^*|S|)$ and
each landmark of $\alpha_2$ within $\alpha$ is decided by only
$\alpha_1$.  This means the long prefix $\alpha_2$ of $\alpha$ is
replaced by the same variables, independent of the occurrence of
$\alpha$.

ESP-comp generates a new shorter string $S^\prime$ of length from $|S|/3$ to
$|S|/2$, and it parses $S^\prime$ iteratively.  Given a string $S$, ESP builds
the ESP-tree of height $O(\lg|S|)$ in $O(|S|\lg^*|S|)$ time and in
$O(|\Sigma\cup V|\lg{|\Sigma\cup V|})$ space.  The approximation ratio of the
smallest grammar by ESP is $O(\lg^*{|S|}\lg{|S|})$~\cite{SakamotoMKS09}.

\subsection{Approximate computations of EDM from ESP-trees}

\begin{figure}[t]
\begin{center}
\includegraphics[width=0.6\textwidth]{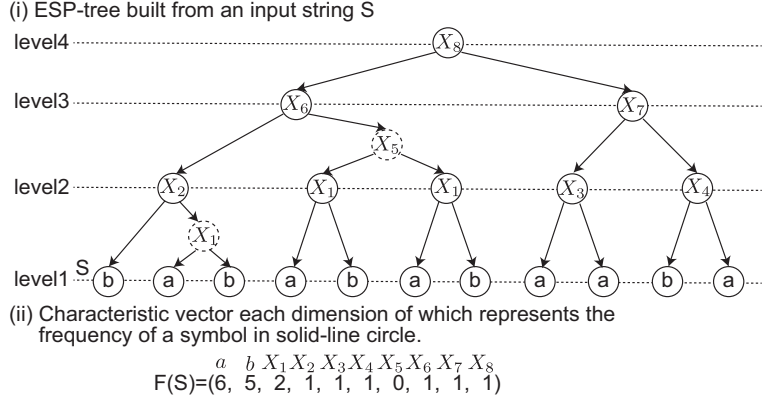}
\end{center}
\vspace{-0.6cm}
\caption{Illustration of ESP-tree and characteristic vector.}
\label{fig:esp}
\end{figure}

ESP-trees enable us to approximately compute EDM for two strings.
After constructing ESP-trees for two strings, 
their \emph{characteristic vectors} are defined as follows.
Let $T(S)$ be the ESP-tree for string $S$.
We define that an integer vector $F(S)$ to be the characteristic vector if
$F(S)(X)$ represents the number of times the variable $X$ appears in $T(S)$ as the root of a 2-tree.
For a string $S$, $T(S)$ and its characteristic vector are illustrated in Figure~\ref{fig:esp}.
The EDM between two strings $S$ and $Q$ can be approximated by $L_1$-distance between two characteristic vectors 
$F(S)$ and $F(Q)$ as follows:
\[
  \|F(S) - F(Q)\|_1 = \sum_{e \in V(S) \cup V(Q)} |F(S)(e) - F(Q)(e)|.
\]

Cormode and Muthukrishnan showed the upper and lower bounds on the
$L_1$-distance between characteristic vectors for the exact EDM.
\begin{Theorem}[\cite{Cormode07}]\label{thm:approx}
For $N=\max(|S|,|Q|)$, 
\[
  d(S,Q) \leq 2\|F(S) - F(Q)\|_1 = O(\lg{N}\lg^*{N})d(S,Q). 
\]
\end{Theorem}

\section{Index Structure for ESP-trees}\label{sec:isESP}
\subsection{Efficient encoding scheme}

\begin{figure}
\includegraphics[width=1.0\textwidth]{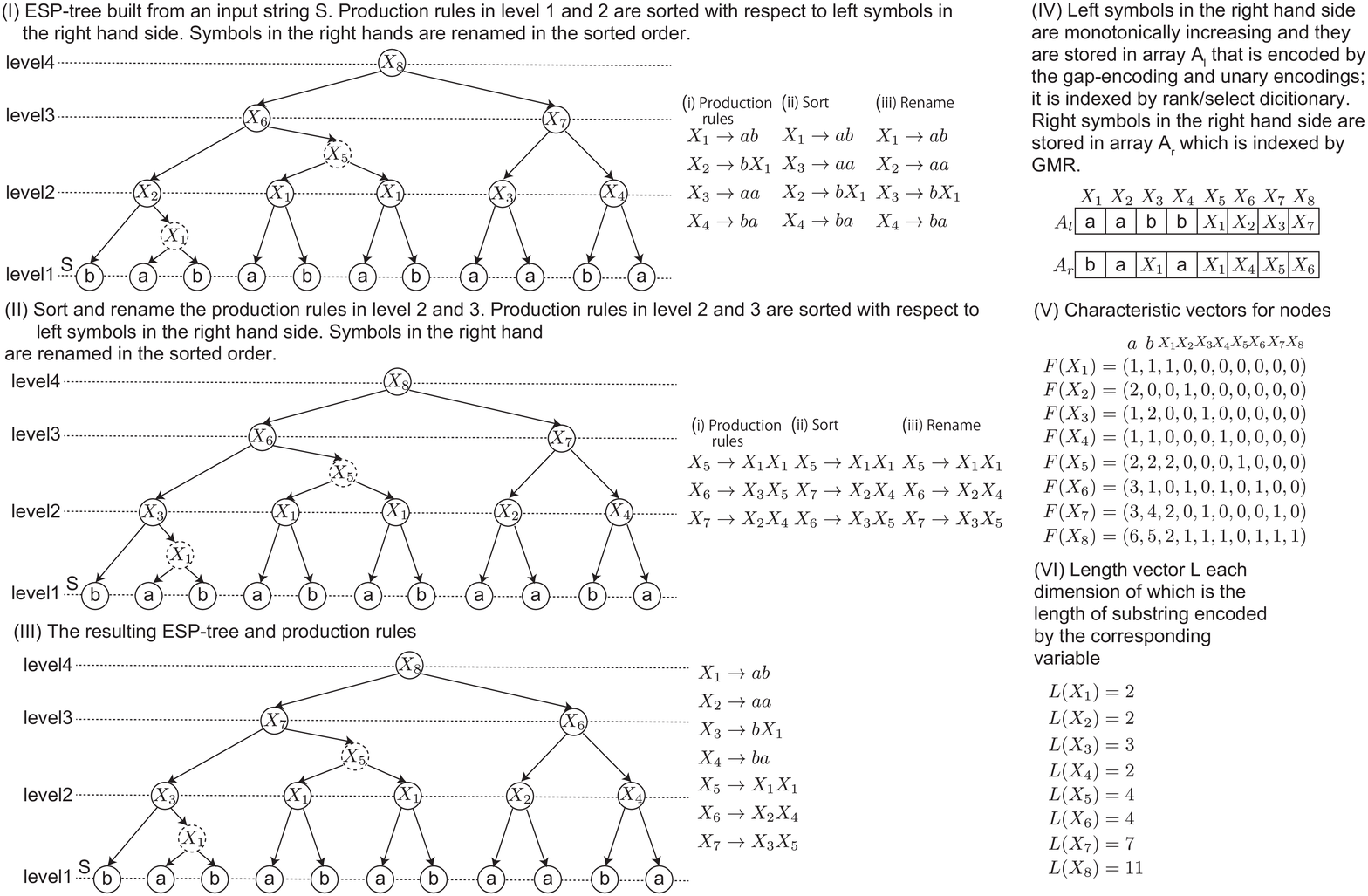}
\caption{Illustration of encoding scheme.}
\label{fig:encoding}
\end{figure}

siEDM encodes an ESP-tree built from a string for fast query searches.
This encoding scheme sorts the production rules in an ESP-tree such
that the left symbols on the right hand side of the production rules
are in monotonically increasing order, which enables encoding of these
production rules efficiently and supporting fast operations for
ESP-trees.  The encoding scheme is performed from the first and second
levels to the top level (i.e., root) in an ESP-tree.

First, the set of production rules at the first and second levels in
the ESP-tree is sorted in increasing order of the left symbols on the
right hand of the production rules, i.e., $X_{l(i)}$ in the form of
$X_i\to X_{l(i)}X_{r(i)}$, which results in a sorted sequence of these
production rules.  The variables in the left hand side in the sorted
production rules are renamed in the sorted order, generating a set of
new production rules that is assigned to the corresponding nodes in
the ESP-tree.  The same scheme is applied to the next level of the
ESP-tree, which iterates until it reaches the root node.

Figure~\ref{fig:encoding} shows an example of the encoding scheme for
the ESP-tree built from an input string $S=babababaaba$.  
At the first and second levels in the ESP-tree, the set of production rules,
$\{X_1\to ab, X_2\to bX_1, X_3\to aa, X_4\to ba\}$, 
is sorted in the lexicographic order of the left symbols on right hand sides of production rules,
which results in the sequence of production rules,
$(X_1\to ab, X_3\to aa, X_2\to bX_1, X_4\to ba)$.  The variables on
the right hand side of the production rules are renamed in the sorted
order, resulting in the new sequence
$(X_1\to ab, X_2\to aa, X_3\to bX_1, X_4\to ba)$, whose production
rules are assigned to the corresponding nodes in the ESP-tree.  This
scheme is repeated until it reaches level~4.

Using the above encoding scheme, we obtain a monotonically increasing
sequence of left symbols on the right hand side of the production
rules, i.e., $X_{l(i)}$ in the form of $X_{i}\to
X_{l(i)}X_{r(i)}$. Let $A_{l}$ be the increasing sequence; $A_{l}$ can
be efficiently encoded into a bit string by using the gap-encoding and
the unary coding.  For example, the gap-encoding represents the
sequence $(1,1,3,5,8)$ by $(1,0,2,2,3)$, and it is further transformed
to the bit string $0^110^010^210^210^31 = 0110010010001$ by unary coding.
Generally, for a sequence $(x_1,x_2,\ldots,x_n)$, its unary code $U$ represents
$x_i$ by $\mathit{rank}_0(U,\mathit{select}_1(U,i))$.
Because the number of $0$s and the number of $1$s is $n +\sigma$ and $n$, respectively, 
the size of $U$ is $2n+\sigma$ bits.  
The bit string is indexed by the rank/select dictionary.

Let $A_r$ be the sequence consisting of the right symbols on the right
hand side of the production rules, i.e., $X_r(i)$ in the form of
$X_i\to X_{l(i)}X_{r(i)}$.  $A_r$ is represented using
$(n+\sigma)\lg{(n+\sigma)}$ bits. $A_r$ is indexed by
GMR~\cite{Golynski06}.

The space for storing $A_l$ and $A_r$ is
$(n+\sigma)\lg{(n+\sigma)}+2n+\sigma+o((n+\sigma)\lg{(n+\sigma)})$ bits in
total.  $A_l$ and $A_r$ enable us to simulate fast queries on encoded
ESP-trees, which is presented in the next subsection.

\subsection{Query processing on tree}

The encoded ESP-trees support the operations
$\mathit{LeftChild}$, $\mathit{RightChild}$, $\mathit{LeftParents}$
and $\mathit{RightParents}$, which are used in our search algorithm.
$\mathit{LeftChild}(X_k)$ returns the left child $X_{l(k)}$ of $X_{k}$
and can be implemented on bit string $A_l$ in $O(1)$ time as
$m=\mathit{select}_1(A_l, X_k)$ and $\mathit{LeftChild}(X_k)=m-X_k$.
$\mathit{RightChild}(X_k)$ returns the right child $X_{r(k)}$ of
$X_{k}$ and can be implemented on array $A_r$ in
$O(\lg\lg{(n + \sigma)})$ time as $X_j=\mathit{access}(A_r,X_k)$.

$\mathit{LeftParents}(X_k)$ and $\mathit{RightParents}(X_k)$ return
sets of parents of $X_k$ as left and right children, respectively,
i.e.,
$\mathit{LeftParents}(X_k)=\{X_i\in V: X_i\to X_kX_j, ^\forall X_j\in
(\Sigma \cup V)\}$ and
$\mathit{RightParents}(X_k)=\{X_i\in V: X_i\to X_jX_k, ^\forall X_j
\in (\Sigma \cup V)\}$.  

Because $A_l$ is a monotonic sequence, any $X_k$ appears consecutively in $A_l$.
Using the unary encoding of $A_l$, $\mathit{LeftParents}(X_k)$ is computed by
$\{p+i : p=\mathit{select}_1(A_l,X_k),\: \mathit{rank}_0(A_l,p+i)=\mathit{rank}_0(A_l,p)\}$ in $O(|\mathit{LeftParents}(X_k)|)$ time.  
$\mathit{RightParents}(X_k)$ can be computed by repeatedly applying select operations for $X_k$ on
$A_r$ until no more $X_k$ appear, i.e., $\mathit{select}_{X_k}(A_r, p)$ for $1\leq p\leq n$.  
Thus, $\mathit{RightParents}(X_k)$ for $X_k\in V$ can be computed in $O(|\mathit{RightParents}(X_k)|)$ time.

\subsection{Other data structures}

As a supplemental data structure, siEDM computes the \emph{node
  characteristic vector}, denoted by $F(X_i)$, for each variable
$X_i$: the characteristic vector consisting of the frequency of any
variable derived from $X_i$.  The space for storing all node
characteristic vectors of $n$ variables is at most $n^2\lg|S|$ bits.
Figure~\ref{fig:encoding}-(V) shows an example of the node
characteristic vectors for ESP-tree in
Figure~\ref{fig:encoding}-(III).  In addition, let $V(X_i)$ be a set
of $X_i$ and variables appearing in all the descendant nodes under
$X_i$, i.e., $V(X_i)=\{e\in (V\cup \Sigma): F(X_i)(e) \neq 0\}$.
Practically, $F(X_i)$ is represented by a sequence of a pair of
$X_j \in V(X_i)$ and $F(X_i)(X_j)$.  Additionally, because
$F(X_i) = F(\mathit{LeftChild}(X_i)) + F(\mathit{RightChild}(X_i)) +
(X_i,1)$ ($+ (X_i,1)$ represents adding $1$ to dimension $X_i$), the
characteristic vectors can be stored per level $2$ of the
ESP-tree. The data structure is represented by a bit array $FB$
indexed by a rank/select dictionary and the characteristic vectors
reduced per level $2$ of ESP-tree. $FB$ is set to $1$ for $i$-th bit
if $F(X_i)$ is stored, otherwise it is $0$.  Then, $F(X_i)$ can be
computed by $\mathit{rank}_1(FB,i)$-th characteristic vector if the $i$-th bit
of $FB$ is $1$; otherwise,
$F(\mathit{LeftChild}(X_i)) + F(\mathit{RightChild}(X_i)) + (X_i,1)$.

Another data structure that siEDM uses is a non-negative integer
vector named \emph{length vector}, each dimension of which is the
length of the substring derived from the corresponding variable (See
Figure~\ref{fig:encoding}-(VI)).  The space for storing length vectors
of $n$ variables is $n\lg{|S|}$ bits.

From the above argument, the space of the siEDM's index structure for
$n$ variables is
$n(n+1)\lg{|S|}+(n+\sigma)\lg{(n+\sigma)}+2n+\sigma+o((n+\sigma)\lg{(n+\sigma)})$
bits in total.

\section{Search Algorithm}\label{sec:sa}

\subsection{Baseline algorithm}

Given an ESP tree $T(S)$, the \emph{maximal subtree decomposition} of $S[i,j]$ 
is a sequence $(X_1,X_2,\ldots,X_m)$ of variable in $T(S)$ defined recursively as follows.
$X_1$ is the variable of the root of the maximal subtree satisfying that
$S[i]$ is its leftmost leaf and $|val(X_1)|\leq j-i$.
If $val(X_1)=S[i,j]$, then $(X_1)$ is the maximal subtree decomposition of $S[i,j]$.
Otherwise, let $X_1,X_2,\ldots,X_m$ be already determined and 
$|val(X_1)val(X_2)\cdots val(X_m)|=k < j-i$.
Then, let $X_{m+1}$ be the variable of the root of the maximal subtree satisfying that
$S[i+k+1]$ is its leftmost leaf and $|val(X_{m+1})|\leq j-i-k$.
Repeating this process until $val(X_1)val(X_2)\cdots val(X_m)=S[i,j]$,
the maximal subtree decomposition is determined.

Based on the maximal subtree decomposition, 
we explain the outline of the baseline algorithm, called online
ESP~\cite{Takabatake14-2}, for computing an approximation of EDM
between two strings.  $T(S)$ is constructed beforehand.  
Given a pattern $Q$, the online ESP computes $T(Q)$, and for each substring
$S[i,j]$ of length $|Q|$, it computes the approximate EDM as follows.
It computes the maximal subtree decomposition $(X_1,X_2,\ldots, X_m)$ of $S[i,j]$. 
Then, the distance $\|F(Q)-F(S[i,j])\|_1$ is approximated by
$\|F(Q) - \sum_{k=1}^mF(X_k)\|_1$ because ESP-tree is balanced and
then $\|F(S[i,j])-\sum_{k=1}^mF(X_k)\|_1 = O(\lg m)$.  This baseline
algorithm is, however, required to compute the characteristic vector
of $S[i,j]$ at each position $i$.  Next, we improve the time and space
of the online ESP by finding those $|Q|$-grams for each variable $X$
in $V(S)$ instead of each position $i$.

\subsection{Improvement}

The siEDM approximately solves Problem~\ref{prob:1} with the same
guarantees presented in Theorem~\ref{thm:approx}.  Let
$X_i \in V(S)$ such that $|\mathit{val}(X_i)| > |Q|$.  There are
$|Q|$-grams formed by the string
$\mathit{suf}(\mathit{val}(X_{l(i)}),
|Q|-k)\mathit{pre}(\mathit{val}(X_{r(i)}), k)$ with
$k=1,2,\ldots,(|Q|-1)$.  Then, the variable $X_i$ is said to
\emph{stab} the $|Q|$-grams.  The set of the $|Q|$-grams stabbed by
$X_i$ is denoted by $\mathit{itv}(X_i)$.  Let $\mathit{itv}(S)$ be the
set of $\mathit{itv}(X_i)$ for all $X_i$ appearing in $T(S)$.  An
important fact is that $\mathit{itv}(S)$ includes any $|Q|$-gram in
$S$.  Using this characteristic, we can reduce the search space .

If a $|Q|$-gram $R$ is in $\mathit{itv}(X_i)$, there exists a maximal
subtree decomposition $X_{i_1},X_{i_2},\ldots, X_{i_m}$.  Then, the
$L_1$-distance of $F(Q)$ and $\sum_{j=1}^{m}{F(X_{i_j})}$ guarantees
the same upper bounds in the original ESP as follows.

\begin{Theorem}\label{thm:newapprox}
  Let $R\in \mathit{itv}(X_i)$ be a $|Q|$-gram on $S$ and
  $X_{i_1},X_{i_2},\ldots, X_{i_m}$ be its maximal subtree
  decomposition in the tree $T(X_i)$.  Then, it holds that
\[
\|F(Q) - \sum_{j=1}^{m}F(X_{i_j})\|_1 = O(\lg{|Q|}\logstar{|S|})d(Q,R).
\]
\end{Theorem}
\begin{Proof}
  By Theorem~\ref{thm:approx},
  $\|F(Q) - F(R)\|_1 = O(\lg{|Q|}\logstar{|S|})d(Q,R)$.  On the other
  hand, for an occurrence of $R$ in $S$, let $T(X_i)$ be the smallest
  subtree in $T(S)$ containing the occurrence of $R$, i.e.,
  $R\in\mathit{itv}(X_i)$.  For $T(R)$ and $T(X_i)$, let $s(R)$ and
  $s(X_i)$ be the sequences of the level 2 symbols in $T(R)$ and
  $T(X_i)$, respectively.  By the definition of the ESP, it holds that
  $s(R)=\alpha\beta\gamma$ and $s(X_i)=\alpha'\beta\gamma'$ for some
  strings satisfying $|\alpha\alpha'\gamma\gamma'|=O(\lg^*|S|)|$, and
  this is true for the remaining string $\beta$ iteratively.  Thus,
  $\|F(R)-F(X_i)\|_1=O(\lg|R|\lg^*|S|)$ since the trees are balanced.
  Hence, by the equation
\begin{align*}
  \|F(Q) - \sum_{j=1}^{m}F(X_{i_j})\|_1 &= O(\lg{|Q|}\logstar{|S|})d(Q,R) + O(\lg{|Q|}\logstar{|S|})\\
                                        &= O(\lg{|Q|}\logstar{|S|})d(Q,R),
\end{align*}
we obtain the approximation ratio.
\end{Proof}
To further enhance the search efficiency, we present a lower bound of the
$L_1$-distance between characteristic vectors, which can be used for reducing
the search space.

\begin{Theorem}[A lower bound $\mu$]\label{thm:lower}
For any $X_i\in V(S)\cup V(Q)$, 
the inequality $\|F(S) - F(Q)\|_1 \geq \mu(X_i)$ holds where
\[
\mu(X_i)=\sum_{e\in V(S) \bigoplus V(Q)}F(X_i)(e).
\]
\end{Theorem}
\begin{Proof}
  The $L_1$ distance between $F(S)$ and $F(Q)$ is divided into four classes of terms: (i) both members in $F(S)$ and $F(Q)$ are non-zero, 
(ii) both members in $F(S)$ and $F(Q)$ are zero, 
(iii) the members in $F(S)$ and $F(Q)$ are zero and non-zero, 
(iv) the members in $F(S)$ and $F(Q)$ are non-zero and zero, respectively.  
Terms consisting of class (iii) and (iv) can be written as $\sum_{e\in V(S) \bigoplus V(Q)}F(S)(e)$, which is a lower bound of the $L_1$-distance.  
Thus, we obtain the inequality $\|F(S) - F(Q)\|_1 \geq \sum_{e\in V(S) \bigoplus V(Q)}F(S)(e)$. 
\end{Proof}

\begin{Theorem}[Monotonicity of $\mu$] \label{thm:mon}
If a variable $X_i$ derives $X_k$, the inequality $\mu(X_i) \geq \mu(X_k)$ holds. 
\end{Theorem}
\begin{Proof}
Every entry in $F(X_k)$ is less than or equal to the corresponding entry in $F(X_i)$. Thus, the inequality holds. 
\end{Proof}

\subsection{Candidate finding}

By Theorems \ref{thm:newapprox}, \ref{thm:lower} and \ref{thm:mon},
the task of the algorithm is reduced to finding a maximal subtree
decomposition $(X_{i_1},X_{i_2},\ldots,X_{i_m})$ within $X_i$.  Given
a threshold $\tau\geq 0$, for each $|Q|$-gram in $\mathit{itv}(S)$,
the algorithm finds the \emph{candidate}: the maximal subtree
decomposition $(X_{i_1},X_{i_2},\ldots,X_{i_m})$ satisfying
$\mu(X_{i_1})+\mu(X_{i_2}) + \cdots + \mu(X_{i_m}) \leq \tau$.

\begin{figure}
\begin{center}
\includegraphics[width=0.63\textwidth]{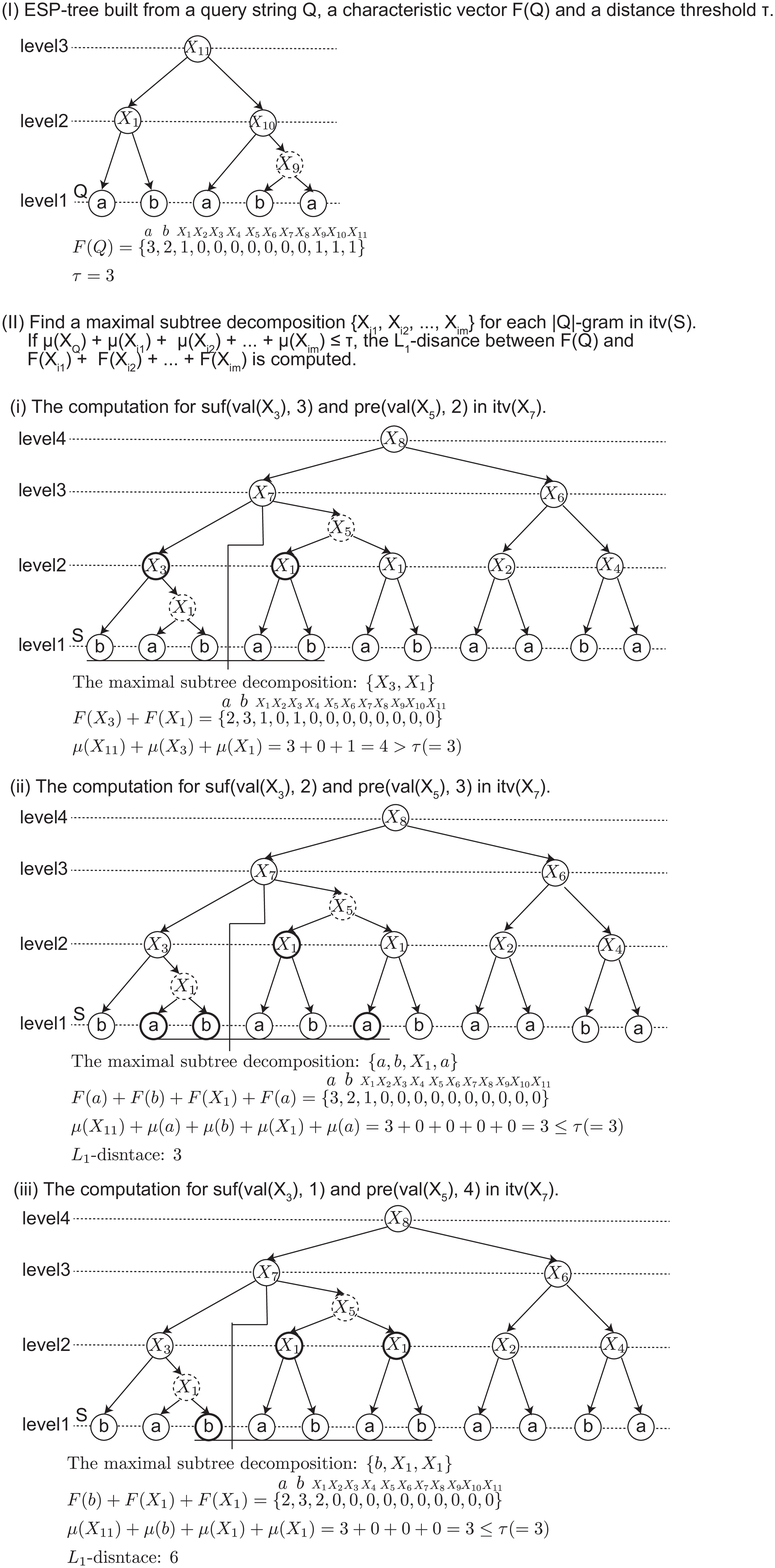}
\caption{Illustration of candidate finding and $L_1$-distance computation.}
\label{fig:search}
\end{center}
\end{figure}

For an $X_i$ and an occurrence of some $|Q|$-gram in $\mathit{itv}(X_i)$,
the $|Q|$-gram is formed by the expression
$\mathit{suf}(\mathit{val}(X_{l(i)}),|Q|-k)\mathit{pre}(\mathit{val}(X_{r(i)}),k)$
for a $k$ $(1\leq k\leq |Q|-1)$.  
The algorithm computes the maximal
subtree decompositions $(x_1,x_2,\ldots,x_p)$ for 
$\mathit{suf}(\mathit{val}(X_{l(i)}),|Q|-k)$ and
$(y_1,y_2,\ldots,y_q)$ for
$\mathit{pre}(\mathit{val}(X_{r(i)}),k)$, and outputs
$(x_1,\ldots,x_p,y_1,\ldots,y_q)$ covering the $|Q|$-gram when
$\sum_{1\leq i\leq p}\mu(x_i)+\sum_{1\leq i\leq q}\mu(y_i)\leq \tau$.
We illustrate the computation of candidates satisfying 
$\mu(X_{i_1})+\mu(X_{i_2}) + \cdots + \mu(X_{i_m}) \leq \tau$ in Figure~\ref{fig:search}
and show the pseudo-code in Algorithm~\ref{alg:search}.

Applying all variables to Algorithm~\ref{alg:search} enables us to
find the candidates covering all solutions.  There are no
possibilities for missing any $|Q|$-grams in $\mathit{itv}(S)$ such
that the $L_1$-distances between their characteristic vectors and
$F(Q)$ are at most $\tau$, i.e., false negatives.  The set may include
a false positive, i.e., the solution set encodes a $|Q|$-gram such
that the $L_1$-distance between its characteristic vector and $F(Q)$
is more than $\tau$.  However, false positives are efficiently removed
by computing the $L_1$-distance
$\|F(Q)-\sum_{j=1}^{m} F(X_{i_j})\|_1$ as a post-processing.

\begin{Theorem}\label{thm:fc}
The time of {\sc FindCandidates} is $O(n|Q|\lg\lg{(n+\sigma)}(\lg{|S|}+\lg{|Q|}))$.
\end{Theorem}
\begin{Proof}
Because the height of the ESP-tree is $O(\lg{|S|})$, 
for each variable $X$, the number of visited nodes is $O(\lg{|Q|}+\lg|S|)$.
The computation time of $\mathit{LeftChild}(X)$ and $\mathit{RightChild}(X)$ is $O(\lg\lg{(n+\sigma)})$, and the
time of {\sc FindLeft} and {\sc FindRight} is $O(|Q|\lg\lg{(n+\sigma)}(\lg{|S|}+\lg{|Q|}))$.
Thus, for $n$ iterations of the functions, the total computation time is $O(n|Q|\lg\lg{(n+\sigma)}(\lg{|S|}+\lg{|Q|}))$.
\end{Proof}

\begin{algorithm}
  \caption{ to output the candidate $R\subseteq V(S)$ for $X\in V(S)$,
    a query pattern $Q$ and a distance threshold $\tau$.  }
\label{alg:search}
{\footnotesize 
\begin{algorithmic}[1]
\Function{FindCandidates}{$X$,$Q$,$\tau$}
\For{$j=1,2,\ldots,|Q|$}
\State $R \leftarrow \phi$ \Comment{Initialize solution set}
\State $q_1 \leftarrow ${\sc FindLeft}($\mathit{LeftChild}(X),|Q|-j,0,R$) \Comment{for left child}
\State $q_2 \leftarrow ${\sc FindRight}($\mathit{RightChild}(X),j,0,R$) \Comment{for right child}
\If{$q_1=0$ and $q_2=0$}
\State Output $R$
\EndIf
\EndFor
\EndFunction
\Function{FindLeft}{$X,q,d,R$}
\If{$d > \tau$}
\State return $\infty$
\ElsIf{$q=0$}
\State return $0$
\ElsIf{$X\in \Sigma$}
\State $d \leftarrow d + 1$~\mbox{if $X \notin V(Q)$}
\State $R\leftarrow R\cup \{X\}$
\State return $q - 1$
\ElsIf{$|\mathit{val}(X)| \leq q$}
\State $d \leftarrow d + \mu(X)$
\State $R\leftarrow R\cup \{X\}$
\State return $q - |\mathit{val}(X)|$
\EndIf
\State $q^\prime \leftarrow $ {\sc FindLeft}($\mathit{RightChild}(X),q,d,R$)
\If{$q^\prime \neq 0$}
\State return {\sc FindLeft}$(\mathit{LeftChild}(X),q^\prime,d,R)$
\EndIf
\EndFunction
\Function{FindRight}{$X,q,d,R$}
\If{$d > \tau$}
\State return $\infty$
\ElsIf{$q=0$}
\State return $0$
\ElsIf{$X\in \Sigma$}
\State $d \leftarrow d + 1$ \mbox{if $X \notin V(Q)$}
\State $R\leftarrow R\cup \{X\}$
\State return $q - 1$
\ElsIf{$|\mathit{val}(X)| \leq q$}
\State $d \leftarrow d + \mu(X)$
\State $R\leftarrow R\cup \{X\}$
\State return $q - |\mathit{val}(X)|$
\EndIf
\State $q^\prime \leftarrow $ {\sc FindRight}($\mathit{LeftChild}(X),q,d,R$)
\If{$q^\prime \neq 0$}
\State return {\sc FindRight}$(\mathit{RightChild}(X),q^\prime,d,R)$
\EndIf
\EndFunction
\end{algorithmic}
}
\end{algorithm}

\subsection{Computing positions}

\begin{algorithm}
\caption{ to compute the set $P$ of all occurrence of $\mathit{val}(X)$ on $S$ for $X \in V(S)$. 
}
\label{alg:position}
{\footnotesize 
\begin{algorithmic}[1]
\Function{ComputePosition}{$X$}
\State $P \leftarrow \{\emptyset \}$ \Comment{Initialize solution set}
\State {\sc Recursion}($X$, $1$)
\EndFunction
\Function{Recursion}{$X$,$p$}
\If{$X$ is the root node}
\State $P\leftarrow P \cup \{p\}$
\State return
\EndIf
\For{each $X_p \in \mathit{RightParents}(X)$} \Comment{$X$ is the right child of $X_p$}
\State {\sc Recursion}($X_p$,$p+|\mathit{val}(X_p)|-|\mathit{val}(X)|$)
\EndFor
\For{each $X_p \in \mathit{LeftParents}(X)$} \Comment{$X$ is the left child of $X_p$}
\State {\sc Recursion}($X_p$,$p$)
\EndFor
\EndFunction
\end{algorithmic}
}
\end{algorithm}

The algorithm also computes all the positions of $\mathit{val}(X_i)$, denoted by
$P(X_i)=\{p\in \{1,2,\ldots,|S|\}:S[p,p+|\mathit{val}(X_i)|-1]=\mathit(X_i)\}$.  Starting from
$X_i$, the algorithm goes up to the root in the ESP-tree built from $S$.  $p$
is initialized to $0$ at $X_i$.  If $X_{k}$ through the pass from $X_i$ to the
root is the parent with the right child $X_{r(k)}$ on the pass, non-negative
integer $(|\mathit{val}(X_k)|-|\mathit{val}(X_{r(k)})|)$ is added to $p$.  Otherwise, nothing is
added to $p$.  When the algorithm reaches the root, $p$ represents a start
position of $\mathit{val}(X_i)$ on $S$, i.e., $\mathit{val}(X_i)=S[p,p+|\mathit{val}(X_i)|-1]$.  To
compute the set $P(X_i)$, the algorithm starts from $X_i$ and goes up to the
root for each parent in $\mathit{RightParents}(X_i)$ and
$\mathit{LeftParents}(X_i)$, which return sets of parents for $X_i$.
Algorithm~\ref{alg:position} shows the pseudo-code.

\begin{Theorem}\label{thm:pc}
  The computation time of $P(X)$ is $O(\mathit{occ}\lg{|S|})$, where
  $\mathit{occ}$ is the number of occurrences of $X$ in $T(S)$.
\end{Theorem}
\begin{Proof}
Using the index structures of $\mathit{RightParents}(X)$ and $\mathit{LeftParents}(X)$,
we can traverse the path from any node with label $(X)$ to the root of $T(S)$ counting the position.
The length of the path is $O(\lg|S|)$.
\end{Proof}

\begin{Theorem}\label{thm:total}
The search time is $O(n|Q|\lg\lg{(n+\sigma)}(\lg{|S|}+\lg{|Q|}) + occ\lg{|S|})$ 
using the data structure of size $n(n+1)\lg{|S|}+(n+\sigma)\lg{(n+\sigma)}+2n+\sigma+o((n+\sigma)\lg{(n+\sigma)})$ bits.
\end{Theorem}
\begin{Proof}
The time for computing $T(Q)$ and $F(Q)$ is $t_1 = O(|Q|\logstar{|S|})$. 
The time for finding candidates is $t_2 = O(n|Q|\lg\lg{(n+\sigma)}(\lg{|S|}+\lg{|Q|}))$ by Theorem~\ref{thm:fc}. 
The time for computing positions is $O(occ\lg{|S|})$ by Theorem~\ref{thm:pc}.
Thus, the total time is $t_1 + t_2 + t_3 = O(n|Q|\lg\lg{(n+\sigma)}(\lg{|S|}+\lg{|Q|}) + occ\lg{|S|})$.
The size of the data structure is derived by the results in Section~\ref{sec:isESP}. 
\end{Proof}

In Theorem~\ref{thm:total},
$n$ and $occ$ are incomparable because 
$occ > n$ is possible for a highly repetitive string.

\section{Experiments}

\begin{table}
\begin{center}
  \caption{Summary of datasets.}
  \begin{tabular}{|l|c|c|c|}
    \hline
     Dataset        & Length & $|\Sigma|$ & Size (MB) \\ 
\hline
     einstein & $467,626,544$ & $139$ & $446$ \\
     cere     & $461,286,644$ & $5$   & $440$ \\
\hline
  \end{tabular}
\label{tab:dataset}
\end{center}
\end{table}

We evaluated the performance of siEDM on one core of a quad-core Intel
Xeon Processor E5540 (2.53GHz) machine with 144GB memory.  We
implemented siEDM using the rank/select dictionary and GMR in
libcds\footnote{\url{https://github.com/fclaude/libcds}}.  We used two
standard benchmark datasets of einstein and cere from repetitive text
collections in the pizza \& chili
corpus\footnote{\url{http://pizzachili.dcc.uchile.cl/repcorpus.html}},
which is detailed in Table~\ref{tab:dataset}.  As a comparison method,
we used the online pattern matching for EDM called online ESP
(baseline)~\cite{Takabatake14-2} that approximates EDM between a query
$Q$ and substrings of the length of $|Q|$ of each position of an input
text.  We randomly selected $S[i,j]$ as the query pattern $Q$ for each
$|Q|=50, 100, 500, 1000$ and examined the performance.

\begin{table}
\begin{center}
  \caption{
Comparison of the memory consumption for the query search
}
  \begin{tabular}{|l|r|r|}
    \hline
     Dataset                   & {\bf einstein} & {\bf cere} \\ \hline
     siEDM~(MB) & $17.12$ & $254.75$ \\
     baseline~(MB) & $6.98$ & $10.95$ \\  
     \hline
  \end{tabular}
\label{tab:memory}
\end{center}
\end{table}

\begin{table}
\begin{center}
  \caption{
Comparison of the index size and construction time
}
  \begin{tabular}{|l|l|r|r|}
    \hline
     \multicolumn{2}{|c|}{Dataset} & {\bf einstein} & {\bf cere} \\ \hline

                & Encoded ESP-tree~(MB)   & $1.18$ & $19.92$ \\
     Index Size & Characteristic vector $F$~(MB)         & $15.35$ & $227.34$ \\
                & Length vector $L$~(MB)  & $0.59$ & $7.49$ \\
\hline
     \multicolumn{2}{|c|}{Construction time~(sec)} & $117.65$ & $472.21$  \\
\hline
  \end{tabular}
\label{tab:construct}
\end{center}
\end{table}

Table~\ref{tab:memory} shows the memory consumption in the search of the siEDM and baseline.
The memory consumption of siEDM was larger than the baseline for both texts
because the baseline does not have characteristic vectors of each node and length vector. 

Table~\ref{tab:construct} shows the size for each component of the index structure and the time 
for building the index structure on einstein and cere datasets. 
Most of the size of the index structure was consumed by the characteristic vector $F$. 
The index size of cere was much larger than that of einstein. 
The index sizes of cere and einstein were approximately 16 megabytes and 256 megabytes, respectively,
because the number of variables generated from cere was much larger than that generated from einstein. 
The number of variables generated from einstein was $305,098$ and the number of variables generated from cere was $4,512,406$. 
The construction times of the index structures were $118$ seconds for einstein and $472$ seconds for cere. 
The results for constructing the index structures demonstrate the applicability of siEDM to moderately large, repetitive texts.

\begin{figure}[t]
\begin{center}
\begin{tabular}{cc}
\hspace{1.5cm}
\includegraphics[width=0.45\textwidth]{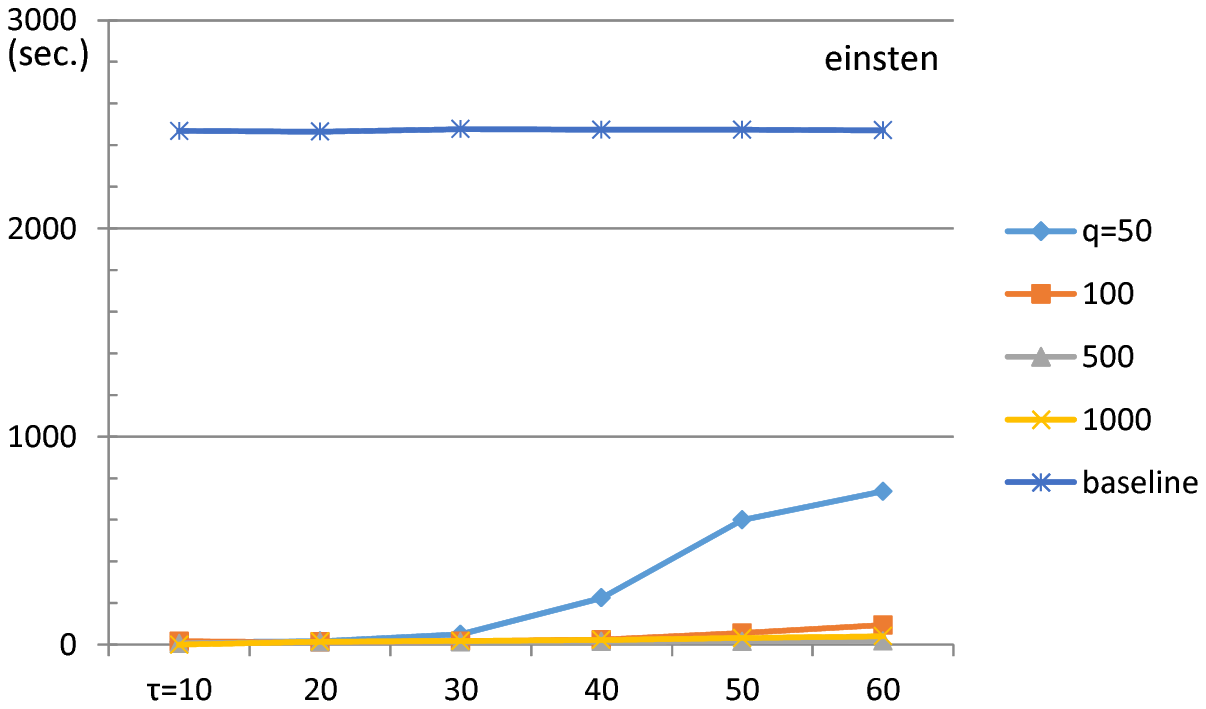} &
\includegraphics[width=0.45\textwidth]{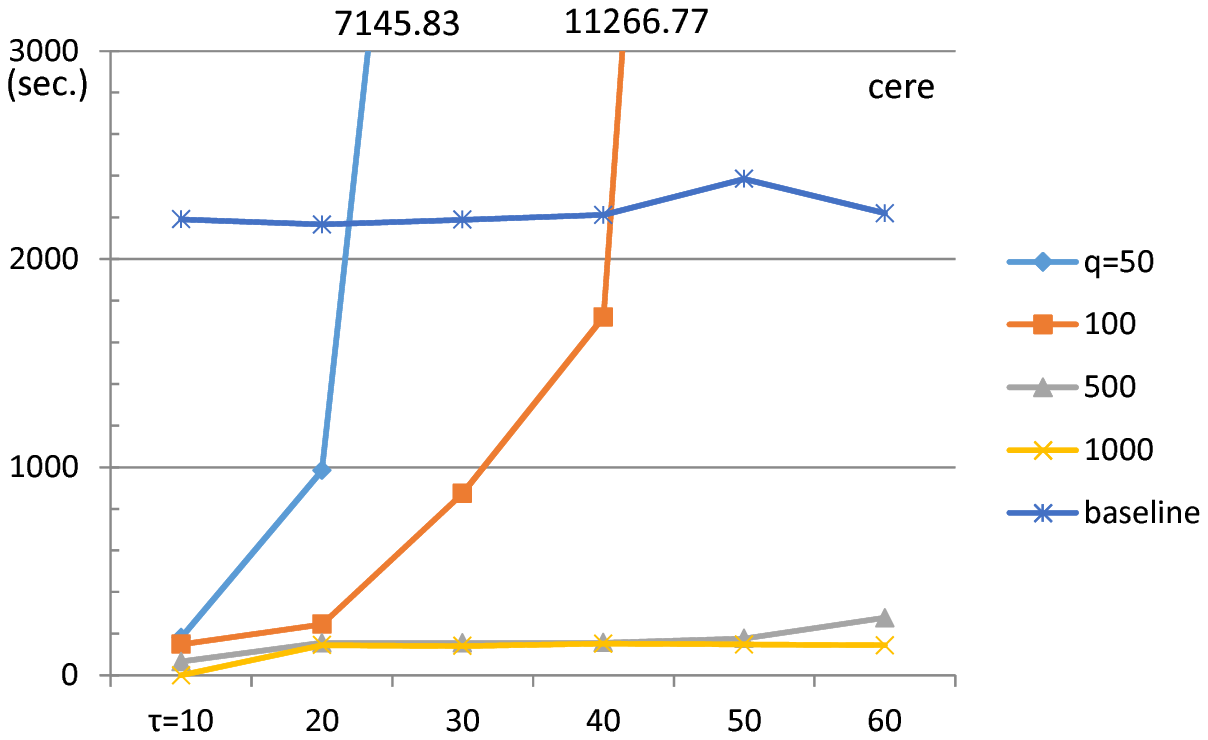} \\\\
\end{tabular}
\end{center}
\caption{Comparison of the search time for einstein (left) and cere (right).}
\label{time}
\end{figure}

\begin{figure}[t]
\begin{center}
\begin{tabular}{cc}
\hspace{1.5cm}
\includegraphics[width=0.45\textwidth]{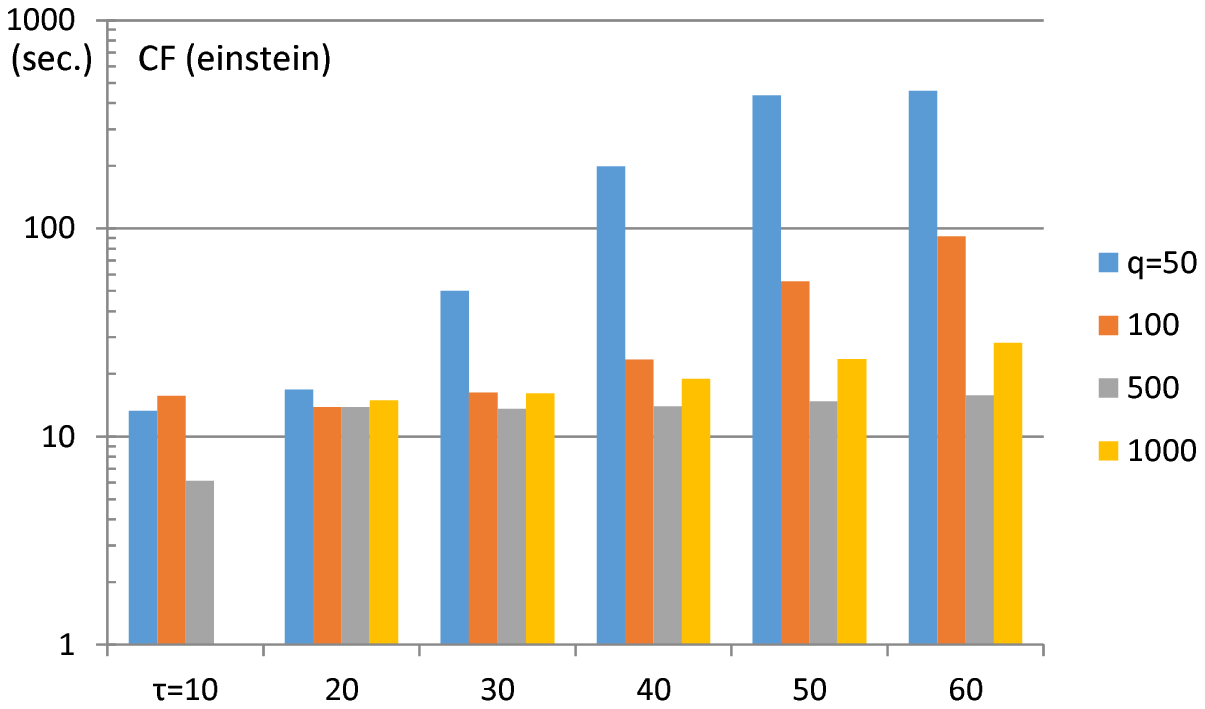} & 
\includegraphics[width=0.45\textwidth]{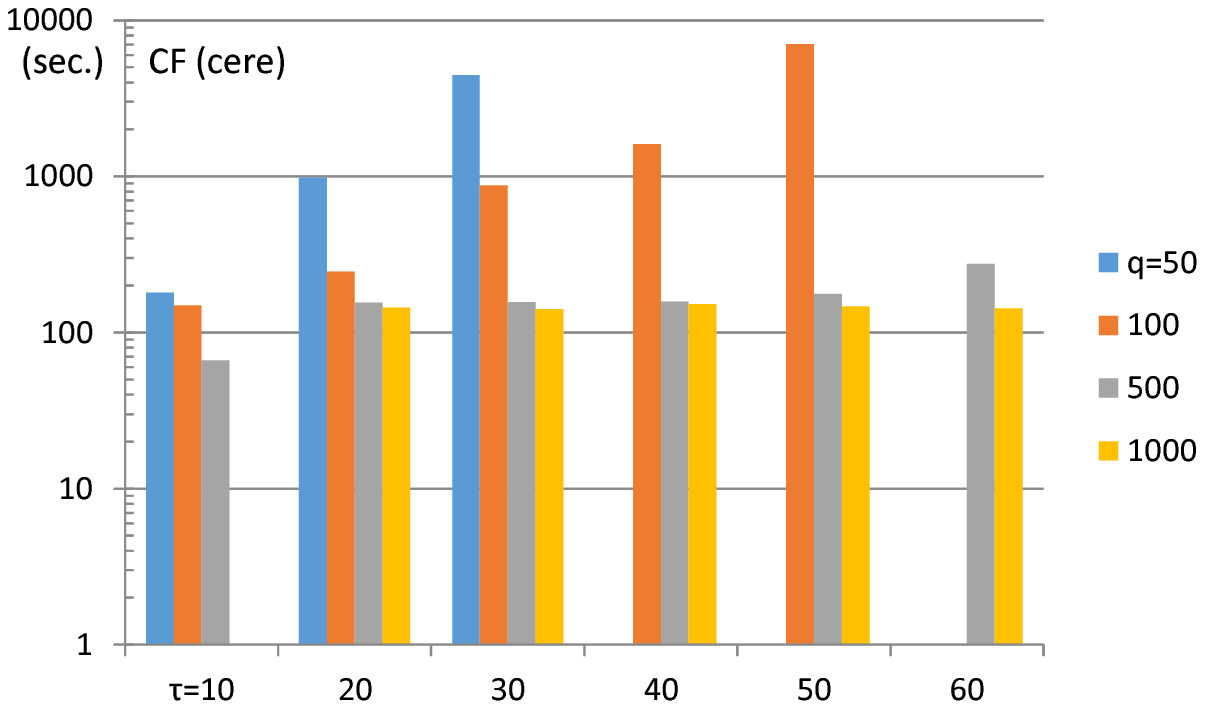} \\
\hspace{1.5cm}
\includegraphics[width=0.45\textwidth]{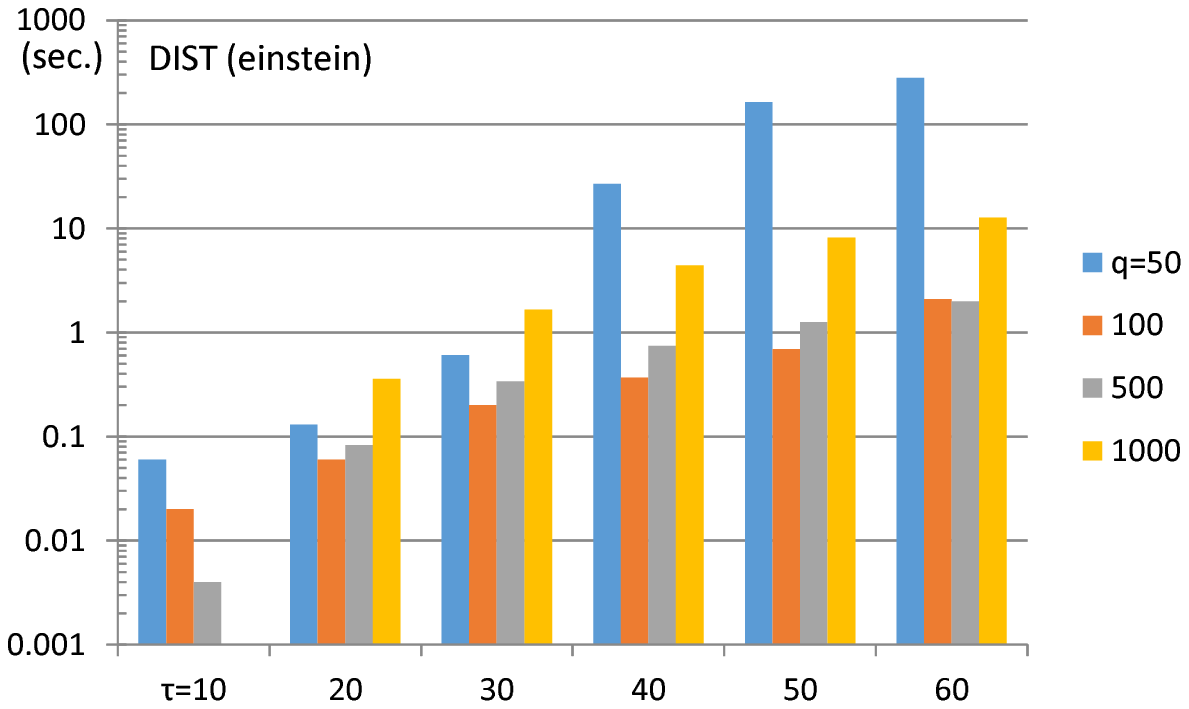} & 
\includegraphics[width=0.45\textwidth]{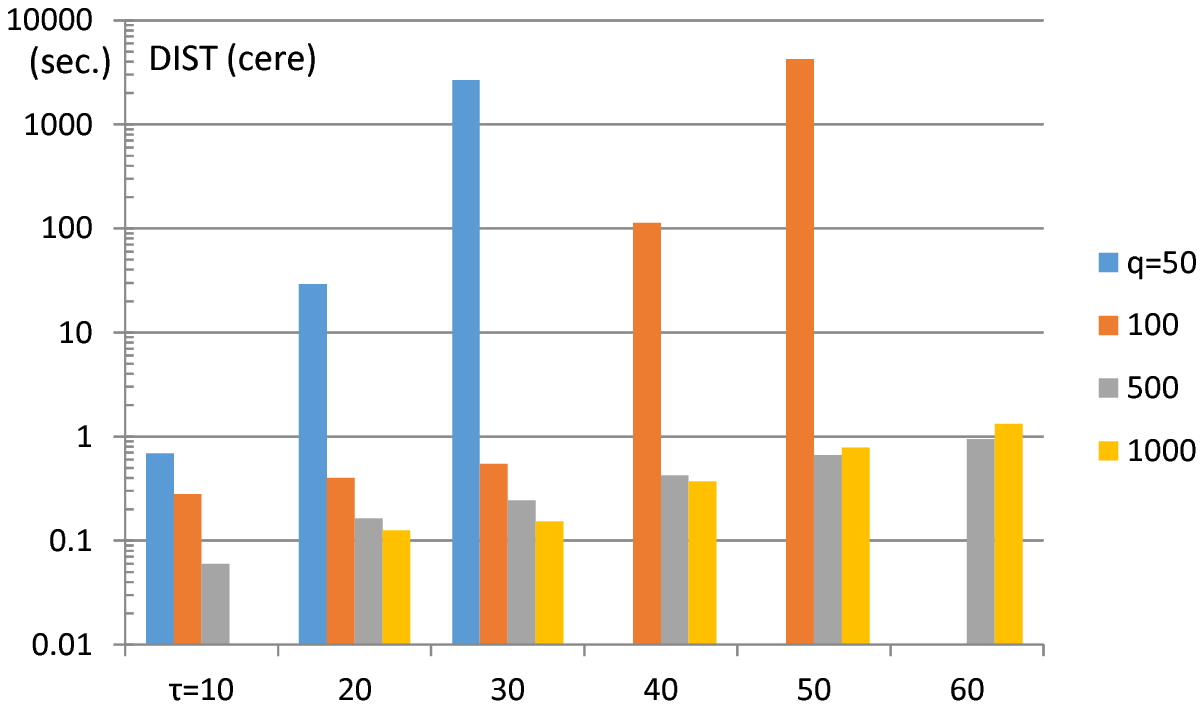} \\
\hspace{1.5cm}
\includegraphics[width=0.45\textwidth]{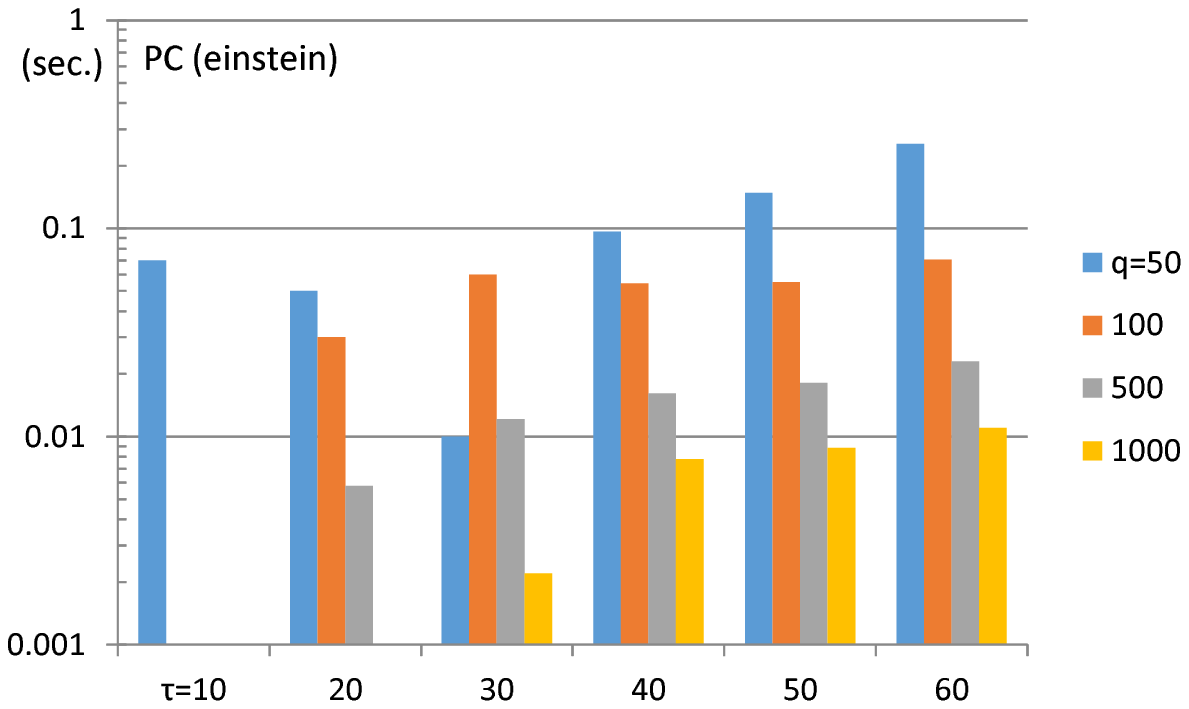} & 
\includegraphics[width=0.45\textwidth]{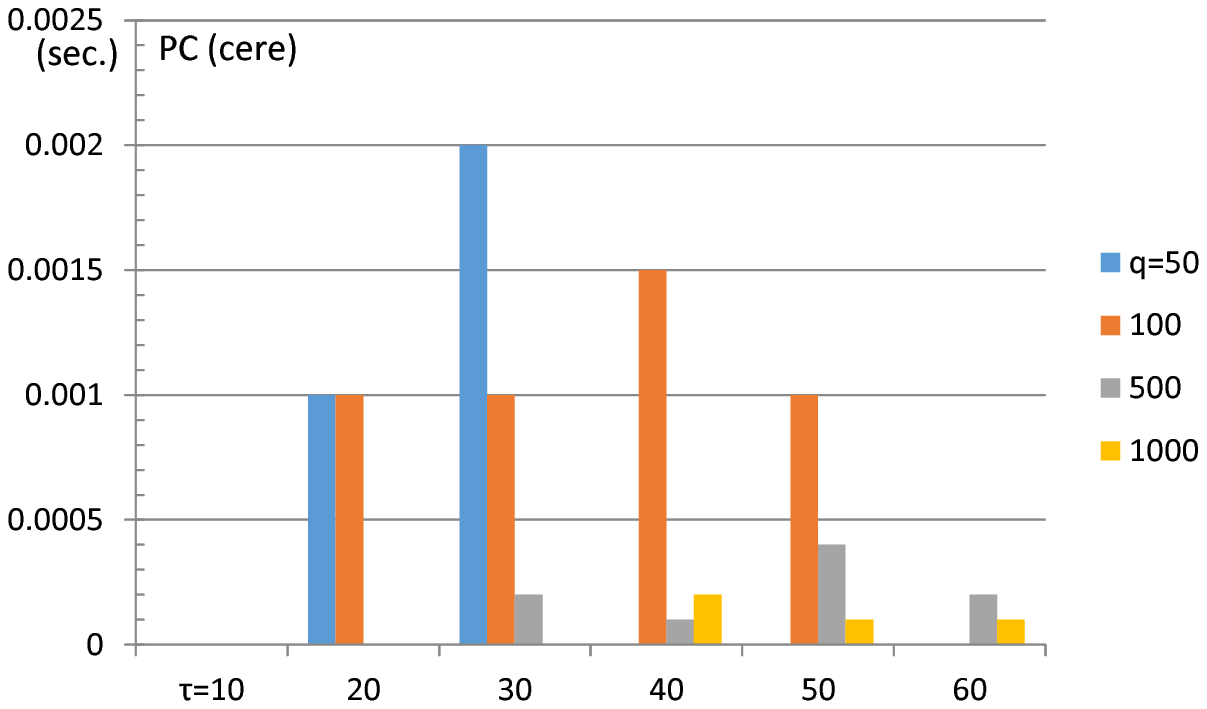} \\\\
\end{tabular}
\end{center}
\vspace{1.5cm}
\caption{Details of search time for different $|Q|$ and $\tau$:
time for candidate findings, CF, time for $L_1$-distance computations, DIST, and
time for position computations, PC.}
\label{time-detail}
\end{figure}

\begin{figure}[t]
\begin{center}
\begin{tabular}{cc}
\hspace{1.8cm}
\includegraphics[width=0.43\textwidth]{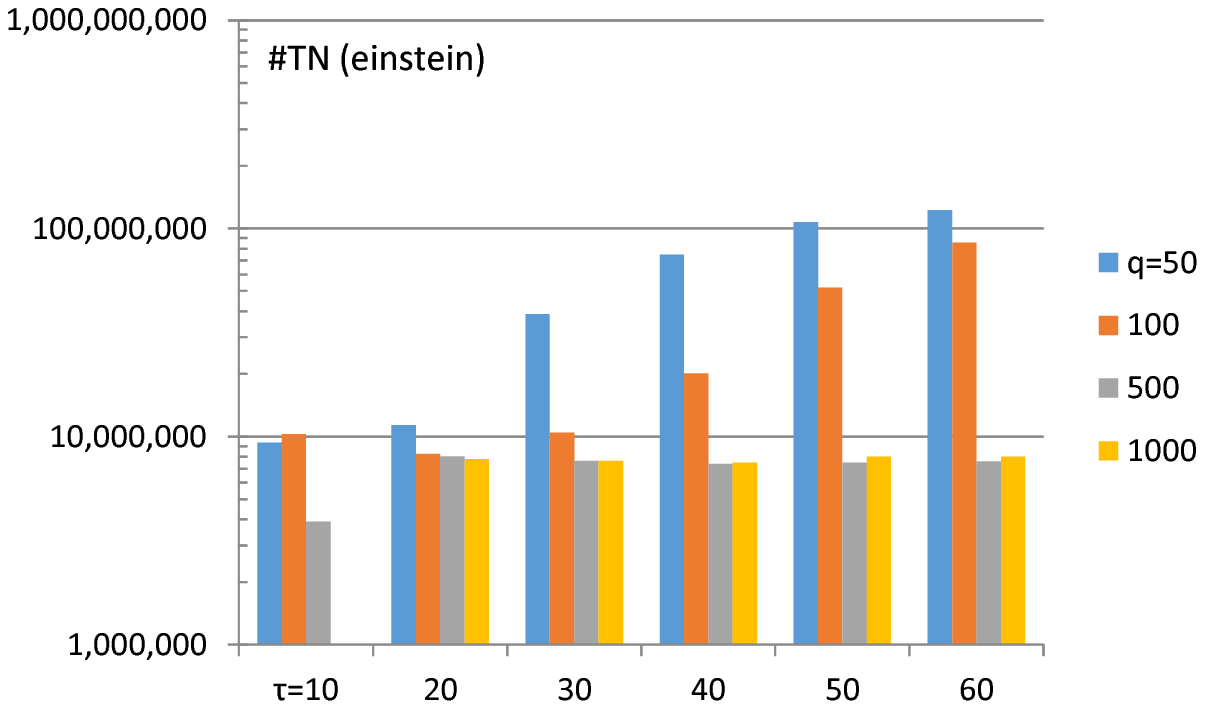} &
\includegraphics[width=0.43\textwidth]{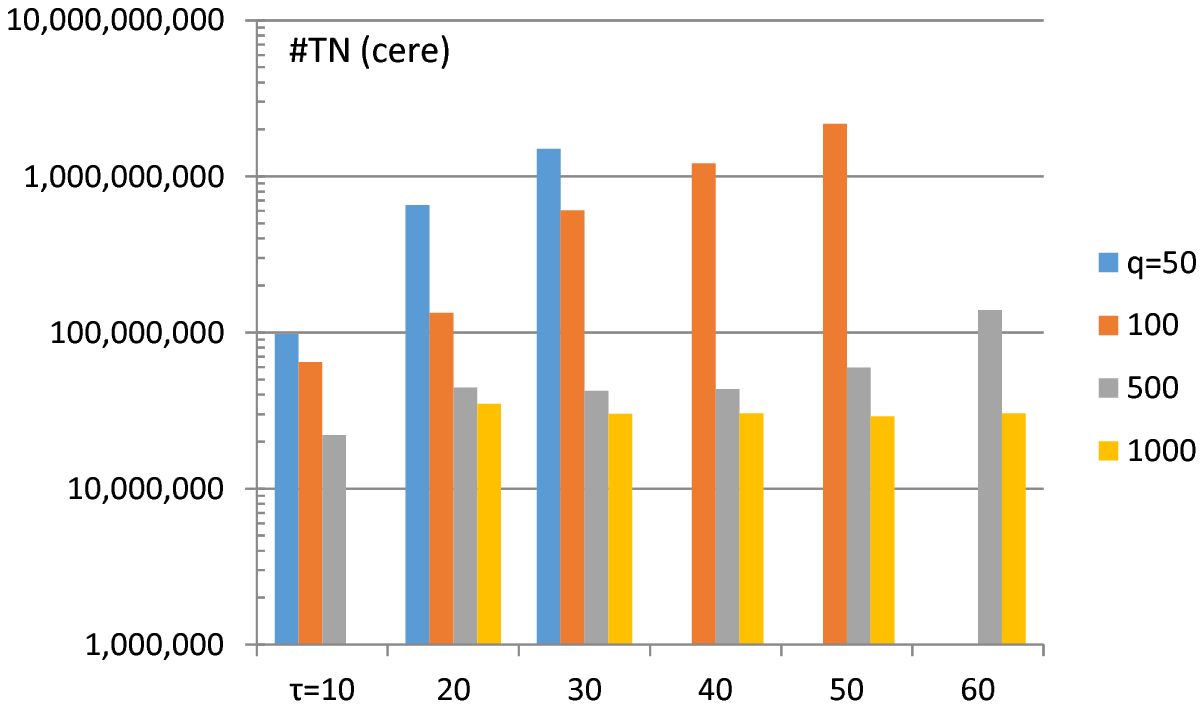} \\
\hspace{1.8cm}
\includegraphics[width=0.43\textwidth]{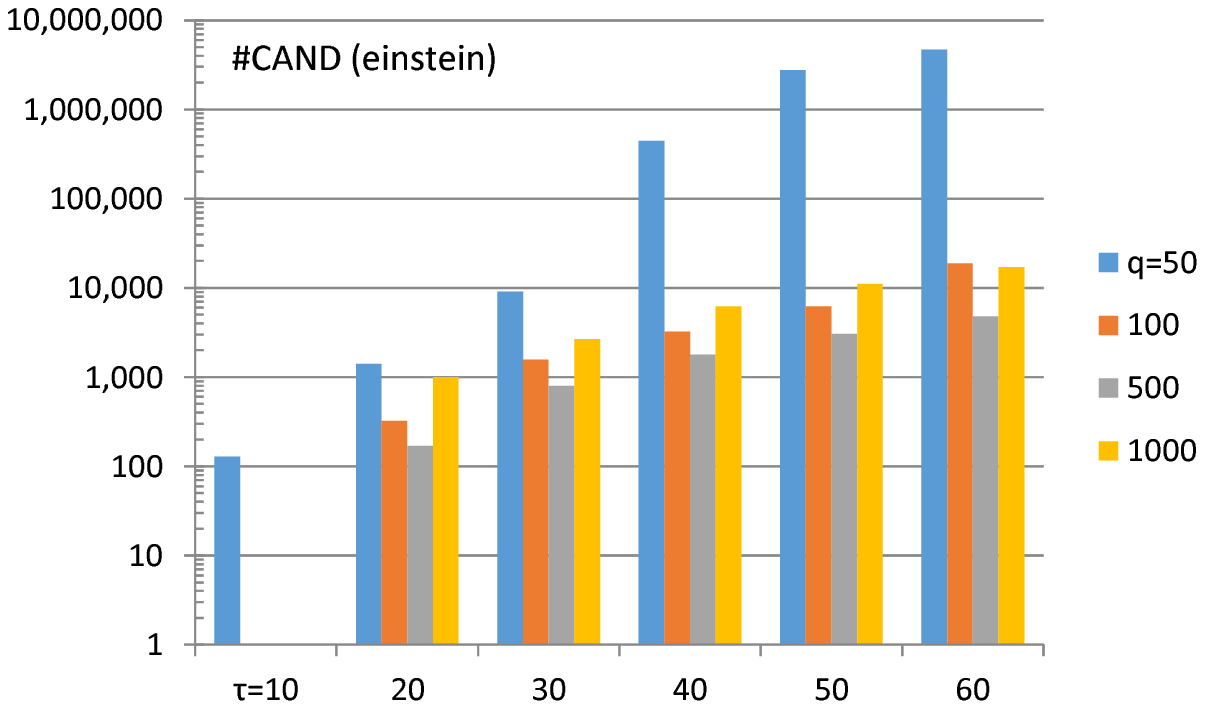} &
\includegraphics[width=0.43\textwidth]{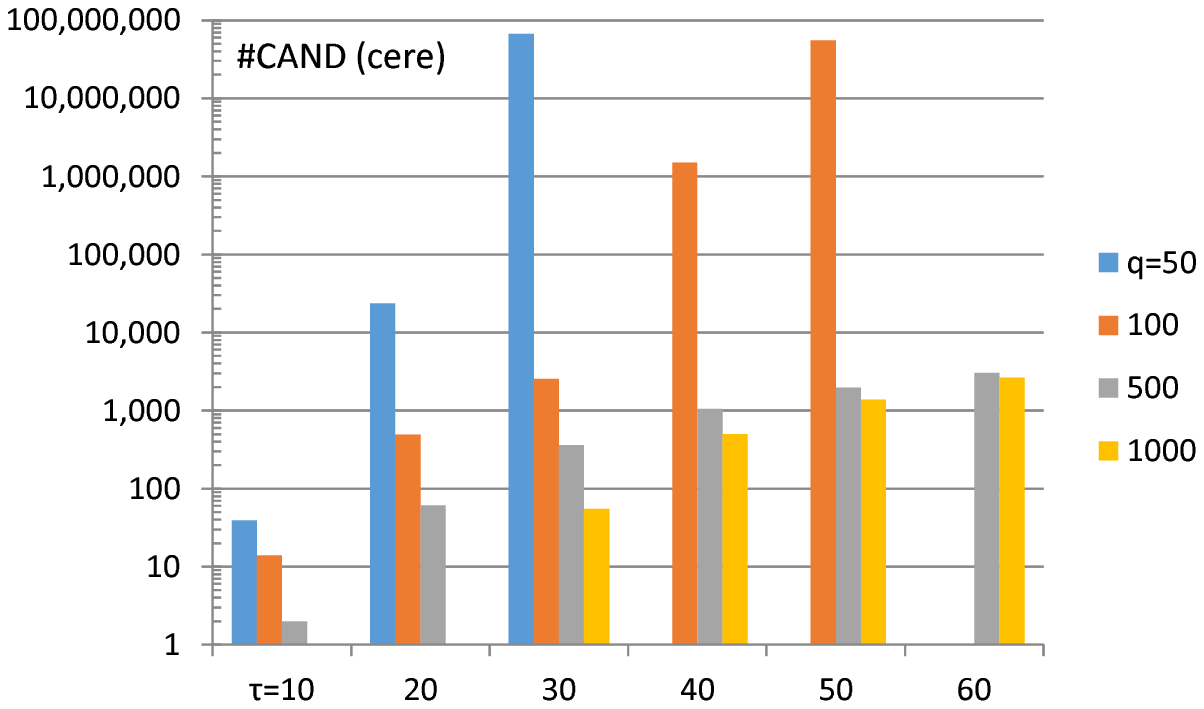} \\
\hspace{1.8cm}
\includegraphics[width=0.43\textwidth]{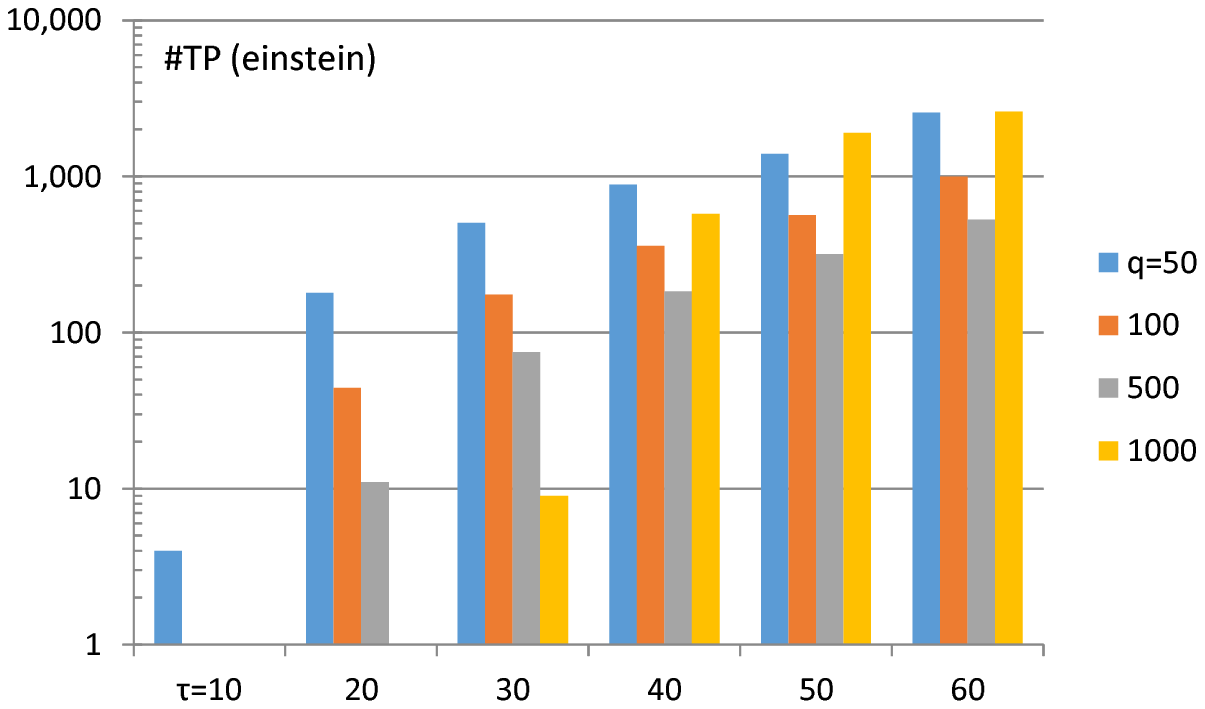} &
\includegraphics[width=0.43\textwidth]{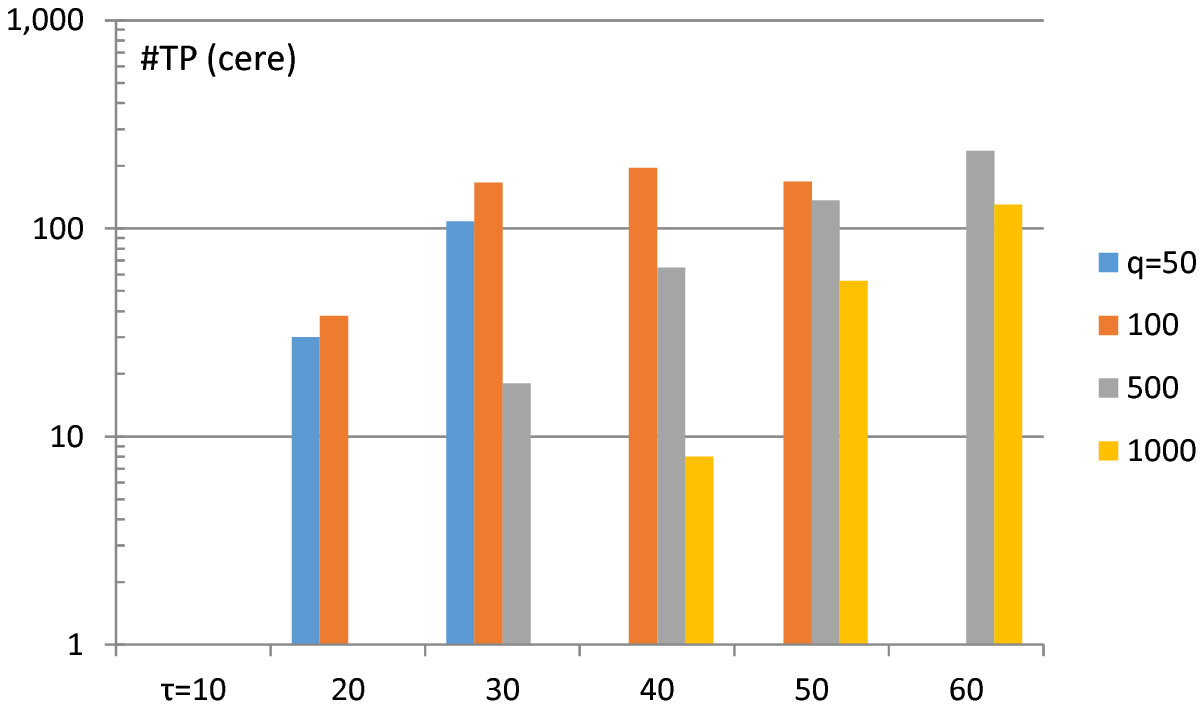} \\
\hspace{1.8cm}
\includegraphics[width=0.43\textwidth]{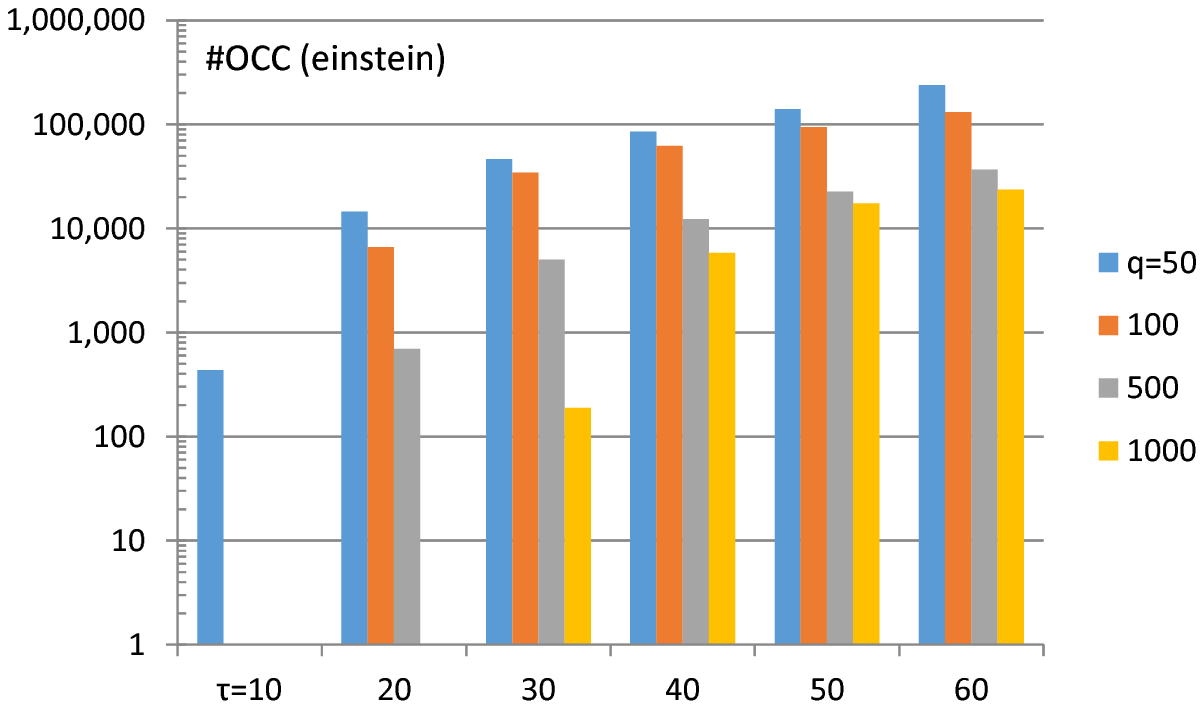} &
\includegraphics[width=0.43\textwidth]{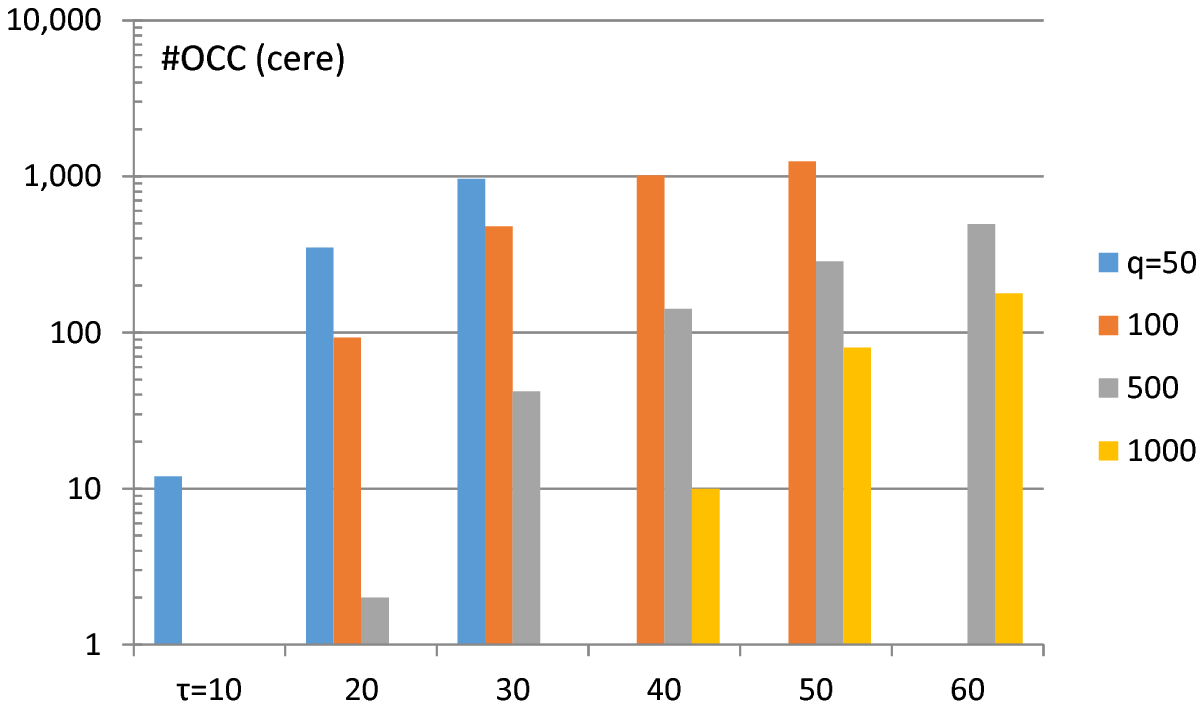} \\\\
\end{tabular}
\end{center}
\caption{Statistical information of the query search: the number of traversed nodes, \#TN,
the number of candidate $|Q|$-grams, \#CAND, 
the number of true positives, \#TP, the number of occurrences, \#OCC.}
\label{statistical}
\end{figure}

Figure~\ref{time} shows the total search time (sec.) of siEDM and the baseline 
for einstein and cere in distance thresholds $\tau$
from $10$ to $60$.
In addition, this result does not contain the case $\tau <10$ because siEDM found no candidate under the condition.
The query length is one of $\{50, 100, 500, 1000\}$. 
Because the search time of baseline is linear in $|S|+|Q|$, we show only the fastest case: $q=|Q|=50$.
The search time of siEDM  was faster than baseline in most cases.

Figure~\ref{time-detail} shows the detailed search time in second.
CF is the time for finding candidates of $Q$ in $T(S)$,
DIST is the time for computing approximated $L_1$ distance by characteristic vectors,
and PC is the time for determining the positions of all $|Q|$-grams within the threshold $\tau$.

Figure~\ref{statistical} shows the number of nodes $T(S)$ visited by the algorithm, \#TN,
the number of candidate $|Q|$-grams computed by ${\sc FindCandidates}$, \#CAND,
the number of true positives among candidate $|Q|$-grams, \#TP, and the number of occurrences, \#OCC.
The most time-consuming task is the candidate finding. 

By the monotonicity of characteristic vectors, 
pruning the search space for small distance thresholds and long query length is more efficient. 
Thus, it is expected that siEDM is faster for smaller distance thresholds and longer query lengths
and the experimental results support this.
The search time on cere is much slower than that on einstein
because the number of generated production rules from cere is much larger than that from einstein,
and a large number of iterations of {\sc FindCandidates} is executed.
In addition, the comparison of \#CAND and \#TP validates the efficiency of siEDM for candidate finding 
with the proposed pruning method.

In Figure~\ref{statistical}, the algorithm failed to find a candidate.
Such a phenomenon often appears when the required threshold $\tau$ is too small, 
because the ESP-tree $T(Q)$ is not necessarily identical to $T(S[i,j])$ even if $Q=S[i,j]$.
Generally, the parsing of $T(S[i,j])$ is affected by a suffix of $S[1,i-1]$ and a prefix of $S[j+1,|S|]$ of length at most $\lg^*|S|$. 

\begin{figure}[t]
\begin{center}
\begin{tabular}{cc}
\hspace{1.5cm}
\includegraphics[width=0.45\textwidth]{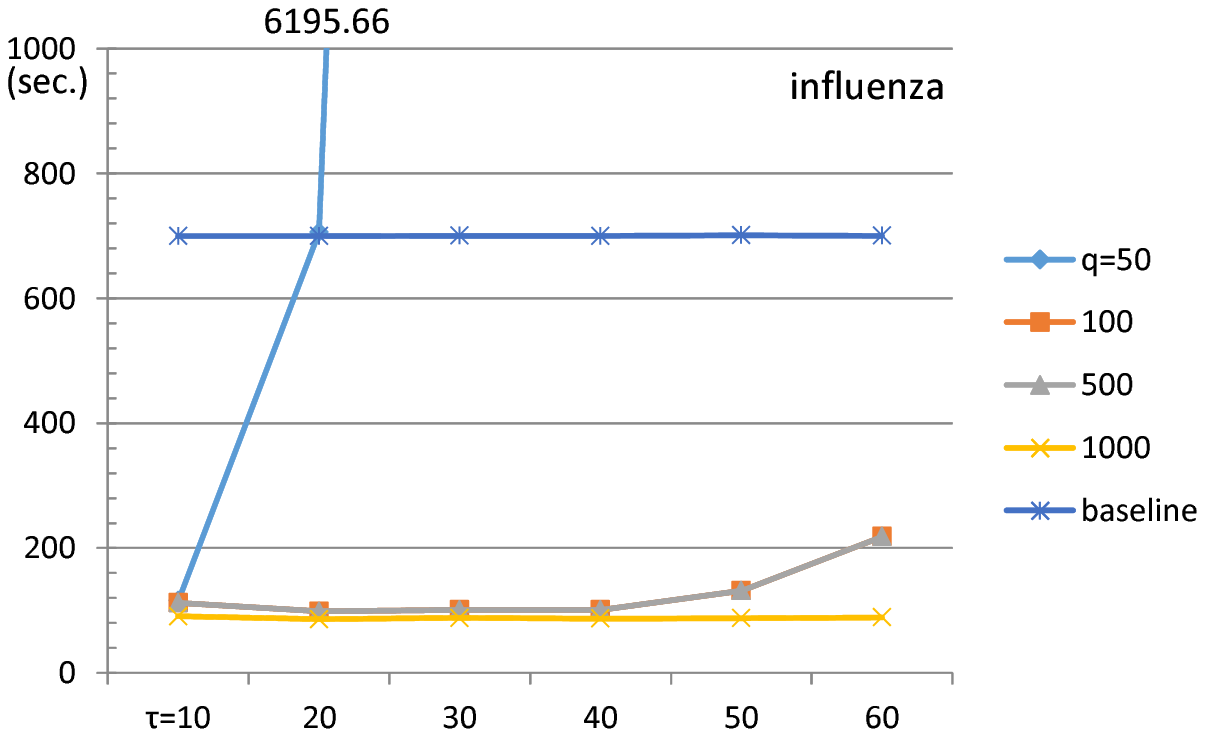} &
\includegraphics[width=0.45\textwidth]{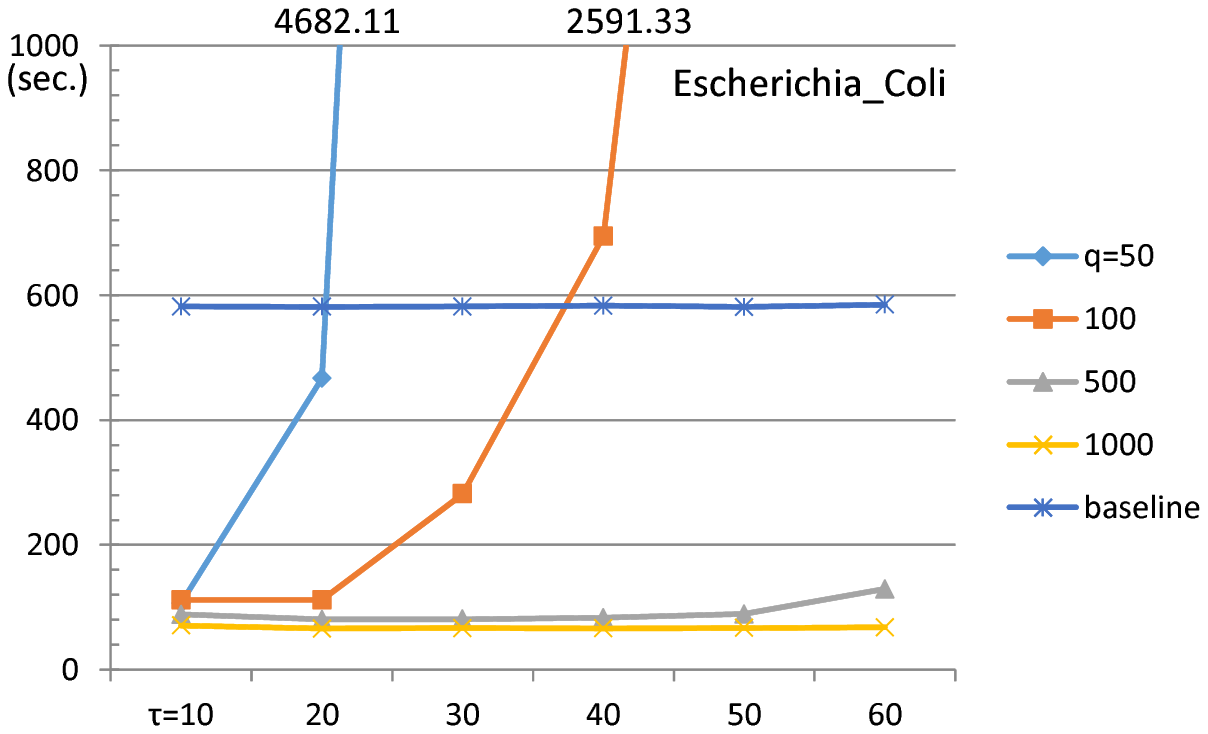}
\end{tabular}
\end{center}
\vspace{1.5cm}
\caption{Search time (sec.) for repetitive texts:
E.Coli (left) and influenza (right).
}
\label{extra}
\end{figure}

As shown in Table~\ref{tab:construct} and Figure~\ref{time},
the search time of siEDM depends on the size of encoded ESP-tree for the input.
Finally, we confirm this feature by an additional experiment for other repetitive texts.
Table~\ref{tab:dataset2}, \ref{tab:memory2} and \ref{tab:construct2} is 
the description of several datasets from the pizza \& chili corpus.
Figure~\ref{extra} shows the search time of siEDM and baseline. 
This result supports our claim that siEDM is suitable for computing EDM of repetitive texts.

\begin{table}[t]
\begin{center}
  \caption{Summary of additional datasets.}
  \begin{tabular}{|l|c|c|c|}
    \hline
     Dataset        & Length & $|\Sigma|$ & Size (MB) \\ 
\hline
     influenza & $154808555$ & $15$ & $147.64$ \\
      Escherichia\_Coli   & $112689515$ & $15$   & $107.47$ \\
\hline
  \end{tabular}
\label{tab:dataset2}
\end{center}
\end{table}

\begin{table}[t]
\begin{center}
  \caption{
Comparison of the memory consumption for the query search
}
  \begin{tabular}{|l|r|r|}
    \hline
     Dataset                   & {\bf influenza} & {\bf Escherichia\_Coli} \\ \hline
     siEDM~(MB) & $164.87$ & $262.01$ \\
     baseline~(MB) & $53.01$ & $100.81$ \\  
     \hline
  \end{tabular}
\label{tab:memory2}
\end{center}
\end{table}

\begin{table}[t]
\begin{center}
  \caption{
Comparison of the index size and construction time for additional datasets.
}
  \begin{tabular}{|l|l|r|r|}
    \hline
     \multicolumn{2}{|c|}{Dataset} & {\bf influenza} & {\bf Escherichia\_Coli} \\ \hline

                & Encoded ESP-tree~(MB)   & $9.92$ & $20.21$ \\
     Index Size & Characteristic vector $F$~(MB)& $150.87$ & $234.91$ \\
                & Length vector $L$~(MB)  & $4.08$ & $6.88$ \\
\hline
     \multicolumn{2}{|c|}{Construction time~(sec)} & $290.33$ & $420.43$  \\
\hline
  \end{tabular}
\label{tab:construct2}
\end{center}
\end{table}

\section{Conclusion}

We have proposed siEDM, an efficient string index for computing approximate
searching based on EDM.  Experimental results demonstrated the applicability
of siEDM to real-world repetitive text collections as well as a longer pattern
search.  Future work will make the search algorithm in siEDM faster, which
would be beneficial for users performing query searches for EDM.

\bibliography{biblio}
\bibliographystyle{plain}

\end{document}